\documentclass{SciPost}

\hypersetup{
    colorlinks,
    linkcolor={red!50!black},
    citecolor={blue!50!black},
    urlcolor={blue!80!black}
}

\usepackage[bitstream-charter]{mathdesign}
\DeclareSymbolFont{usualmathcal}{OMS}{cmsy}{m}{n}
\DeclareSymbolFontAlphabet{\mathcal}{usualmathcal}

\fancypagestyle{SPstyle}{
\fancyhf{}
\lhead{\colorbox{scipostblue}{\bf \color{white} ~SciPost Physics }}
\rhead{{\bf \color{scipostdeepblue} ~Submission }}

\fancyfoot[C]{\textbf{\thepage}}
}

\usepackage{bm}
\usepackage{bbm}
\usepackage{soul} 
\usepackage{mathtools}
\usepackage{orcidlink}
\usepackage{booktabs} 

\newcommand{\beq}{\begin{equation}}
\newcommand{\eeq}{\end{equation}}
\newcommand{\bea}{\begin{eqnarray}}
\newcommand{\eea}{\end{eqnarray}}

\newcommand{\eqlab}[1]{\label{eq:#1}}
\newcommand{\eq}[1]{Eq.~(\ref{eq:#1})}
\newcommand{\Eq}[1]{Equation~(\ref{eq:#1})}
\newcommand{\eqs}[2]{Eqs.~(\ref{eq:#1}) and (\ref{eq:#2})}

\newcommand{\figlab}[1]{\label{fig:#1}}
\newcommand{\fref}[1]{Fig.~\ref{fig:#1}}
\newcommand{\Fref}[1]{Figure~\ref{fig:#1}}

\newcommand{\seclab}[1]{\label{sec:#1}}
\newcommand{\sref}[1]{Sec.~\ref{sec:#1}}
\newcommand{\srefs}[2]{Secs.~\ref{sec:#1} and \ref{sec:#2}} 
\newcommand{\Sref}[1]{Section~\ref{sec:#1}}
\newcommand{\aref}[1]{Appendix~\ref{sec:#1}}

\newcommand{\ket}[1]{\vert#1\rangle}
\newcommand{\bra}[1]{\langle#1\vert}
\newcommand{\ip}[2]{\langle#1\vert#2\rangle}
\newcommand{\me}[3]{\langle#1\vert#2\vert#3\rangle}

\newcommand{\Tr}{\mathrm{Tr}}
\let\ksave\k \renewcommand{\k}{{\bf k}}
\let\rsave\r \renewcommand{\r}{{\bf r}} 
\let\asave\a \renewcommand{\a}{{\bf a}}
\let\bsave\b \renewcommand{\b}{{\bf b}}
\newcommand{\bnk}{_{n\k}}

\newcommand{\blk}{_{l\k}}
\newcommand{\0}{\mathbf{0}}

\newcommand{\w}{\omega}

\newcommand{\Fk}{F_\k}

\newcommand{\enk}{\epsilon\bnk}
\newcommand{\elk}{\epsilon\blk}

\newcommand{\Eg}{E_{\rm g}}

\newcommand{\mm}{\mathfrak{m}}

\newcommand{\one}{\mathbbm{1}}

\newcommand{\al}{\alpha}
\newcommand{\be}{\beta}
\newcommand{\ab}{{\alpha\beta}}

\newcommand{\tr}{\mathrm{tr}}

\renewcommand{\Re}{\mathrm{Re}}
\renewcommand{\Im}{\mathrm{Im}}
\begin{document}

\pagestyle{SPstyle}

\begin{center}{\Large \textbf{\color{scipostdeepblue}{Electronic
      bounds in magnetic crystals }}}\end{center}

\begin{center}
\textbf{Daniel Passos\,\orcidlink{0000-0003-4825-3325}\,,\textsuperscript{1}
  and Ivo Souza\,\orcidlink{0000-0001-9901-5058}\,,\textsuperscript{1,2} }
\end{center}

\begin{center}
{\bf 1} Centro de F{\'i}sica de Materiales, Universidad del País Vasco, 20018 San Sebasti{\'a}n,
  Spain
\\
{\bf 2} Ikerbasque Foundation, 48013 Bilbao, Spain
\end{center}

\section*{\color{scipostdeepblue}{Abstract}}
\textbf{\boldmath{We present a systematic study of bound relations
    between different electronic properties of magnetic crystals:
    electron density, effective mass, orbital magnetization,
    localization length, Chern invariant, and electric
    susceptibility. All relations are satisfied for a group of
    low-lying bands, while some remain valid for upper bands.  New
    results include a lower bound on the electric susceptibility of
    Chern insulators, and an upper bound on the sum-rule part of the
    orbital magnetization.  In addition, bounds involving the Chern
    invariant are generalized from two dimensions (Chern number) to
    three (Chern vector). Bound relations are established for metals
    as well as insulators, and are illustrated for model systems. The
    manner in which they approach saturation in a model Chern
    insulator with tunable flat bands is analyzed in terms of the
    optical absorption spectrum.}}

\vspace{\baselineskip}

\noindent\textcolor{white!90!black}{%
\fbox{\parbox{0.975\linewidth}{%
\textcolor{white!40!black}{\begin{tabular}{lr}%
  \begin{minipage}{0.6\textwidth}%
    {\small Copyright attribution to authors. \newline
    This work is a submission to SciPost Physics. \newline
    License information to appear upon publication. \newline
    Publication information to appear upon publication.}
  \end{minipage} & \begin{minipage}{0.4\textwidth}
    {\small Received Date \newline Accepted Date \newline Published Date}%
  \end{minipage}
\end{tabular}}
}}
}


\vspace{10pt}
\noindent\rule{\textwidth}{1pt}
\tableofcontents
\noindent\rule{\textwidth}{1pt}
\vspace{10pt}


\section{Introduction}
\seclab{intro}

The quantum geometry of electrons in crystals is central to modern
condensed-matter physics.  Its basic ingredients --~Berry phase, Berry
curvature and quantum metric~\cite{provost80,berry84}~-- describe
fundamental ground-state and linear-response
properties~\cite{xiao-rmp10,resta-epjb11,vanderbilt-book18}: electron
polarization and localization, orbital magnetization, and anomalous
Hall conductivity; the quantized anomalous Hall conductivity of Chern
insulators is the prototypical topological response.  New platforms
such as moir\'e materials provide ample opportunities for studying
quantum-geometric electronic properties, and are partly responsible
for the current interest in
them~\cite{yu-npjquantummater25,verma2025quantumgeometryrevisitingelectronic,Gao2025quantumgeometry}.

Quantum-geometric quantities satisfy rigorous bound relations, e.g.,
the metric-curvature
inequalities~\cite{roy-prb14,peotta-natcomms15,ozawa-prb21}.  Such
relations follow from the positive-semidefiniteness of the underlying
tensor objects, and some can be deduced from optical sum rules.  In
insulators, additional sum-rule inequalities, with counterparts in
atomic physics~\cite{traini-epj96}, connect many-electron
localization, electron density, band gap, and electric
susceptibility~\cite{souza-prb00,aebischer-prl01,martin-book04,onishi-prx24,komissarov-natcomms24,onishi-prb24,verma-pnas25,souza-scipost25,onishi-prr25}.
Combining different inequalities allows for example to place upper
bounds on the energy gap of Chern insulators~\cite{onishi-prx24}, and
to bracket the localization length, which can then be estimated from
tabulated parameters~\cite{souza-scipost25}.  The manner in which
various bounds become tight in flat-band systems has been a topic of
continued interest, after the saturation of the metric-curvature
inequality was identified as a criterion for realizing fractional
quantum Hall states~\cite{roy-prb14}.

Metric-curvature and related inequalities have been studied mostly in
two-dimensional (2D) gapped systems: the integer quantum-Hall state of
Landau levels~\cite{ozawa-prb21}, tight-binding models of Chern
insulators~\cite{roy-prb14,ozawa-prb21}, and continuum models of
moir\'e materials~\cite{onishi-prx24,komissarov-natcomms24}.
Extensions to higher-dimensional Chern insulators have been
suggested~\cite{ozawa-prb21,onishi-prx24} but not explicitly worked
out, and applications to metals have also not been much explored.

In this work, we develop bound relations for 2D and 3D magnetic
crystals, both insulating and metallic.  To present them in an unified
manner, we employ generalized Chern coefficients --~Chern number in 2D
and Chern vector in 3D~-- that reduce to the Chern invariants for
insulators, becoming nonquantized in ferromagnetic metals. We find,
for example, that the magnitudes of the Chern coefficients place upper
bounds on the minimum direct gap of 2D and 3D Chern insulators and
ferromagnetic metals; in the case of Chern insulators, the magnitude
squared of the Chern invariant also places a lower bound on the
electric susceptibility.  The former result reduces to a known
relation for 2D Chern insulators~\cite{onishi-prx24}; the latter
recovers as a special case the proportionality between susceptibility
and Chern number in the integer quantum-Hall state of a 2D
free-electron gas in a transverse magnetic
field~\cite{komissarov-natcomms24}.

We also discuss upper bounds on the bulk orbital magnetization
--~established in a recent work~\cite{shinada-prb25}~-- and on its
sum-rule part~\cite{kunes-prb00,souza-prb08}, the intrinsic orbital
magnetic moment of the filled states, elucidating the conditions under
which those bounds saturate.  An example of a saturated orbital-moment
bound is the effective Bohr magneton expression for the valley orbital
moment in the low-energy description of graphene with broken inversion
symmetry~\cite{xiao-prl07}.

2D relations are illustrated for Landau levels, and for two different
tight-binding models with quantum anomalous Hall phases: the Haldane
model~\cite{haldane-prl88}, and a three-band model with tunable band
flatness~\cite{yang-prb12}. To illustrate 3D relations, we consider a
layered Haldane model with trivial and Chern gapped phases separated
by a gapless Weyl-semimetal phase~\cite{liu-nature22}.

The paper is organized as follows. \Sref{defs} establishes basic
definitions and notation, and states the relevant optical sum
rules. \Sref{survey} introduces the metric-curvature tensor and two
related objects, denoted as the mass-moment and mass-magnetization
tensors. They are all positive semidefinite, and some general
implications of this property are discussed in \sref{posdef}.  In
subsequent sections, a systematic study of bound relations arising
from positive semidefiniteness is carried out.  \Sref{roy} addresses
matrix-invariant inequalities in 2D (including the metric-curvature
inequalities), and their generalization to 3D is the subject of
\sref{roy3d}.  \Sref{spectral} considers sum-rule inequalities
involving the energy gap and the electric susceptibility; these are
then combined with matrix-invariant inequalities to place bounds on
the direct gap of Chern insulators and ferromagnetic metals, and on
the electric susceptibility of Chern insulators. Illustrative examples
are presented throughout Secs.~\ref{sec:roy}--\ref{sec:spectral}.  The
paper concludes in \sref{conclusions} with a summary, followed by four
appendices with derivations and accessory results.

\section{Definitions, notation and sum rules}
\seclab{defs}

We treat electrons in crystals in the mean-field approximation.  The
spectral problem is
\beq
H_\k \ket{u\bnk} = \enk\ket{u\bnk}\,,
\eeq
with $\ket{u\bnk}$ the cell-periodic part of a Bloch eigenfunction and
$H_\k = e^{-i\k\cdot\r} H e^{i\k\cdot\r}$.  $H$ may be either a
low-energy effective Hamiltonian, or a microscopic Hamiltonian of the
form~\cite{onishi-prx24}
\beq
H=\frac{p^2}{2m_{\rm e}} + V(\r) +
{\bf p}\cdot {\bf A}(\r) +
{\bf A}(\r)\cdot{\bf p}
\,,
\eqlab{H-loc}
\eeq
where $V(\r)$ and ${\bf A}(\r)$ are lattice-periodic functions.  Upon
replacing the bare electron mass $m_{\rm e}$ with an effective mass,
\eq{H-loc} also applies to certain low-energy
models~\cite{onishi-prx24}.

Given a subset $\Fk$ of Bloch eigenstates at $\k$ (for example, the
filled states at zero temperature), we associate with it a series of
Cartesian tensors labeled by $p\in\mathbbm{Z}$,
\begin{subequations}
\begin{align}
T^\ab_p(\k) &= \sum_{n\in\Fk}\,
\me{\partial_\al u\bnk}{\left( \one - P_\k \right)
  \left( H_\k - \enk \right)^p \left( \one - P_\k \right)}
{\partial_\be u\bnk}
\eqlab{T-geom}
\\
&=\sum_{\substack{n\in\Fk\\{l\notin \Fk}}}
(\elk-\enk)^p r^\alpha_{nl\k} r^\beta_{ln\k}
\eqlab{T-sos1}
\\
&=
\sum_{\substack{n\in\Fk\\{l\notin \Fk}}}
\frac{\me{u\bnk}{(\partial_\al H_\k)}{u\blk}
  \me{u\blk}{(\partial_\be H_\k)}{u\bnk}
}{(\elk-\enk)^{2-p}}\,.
\eqlab{T-sos2}
\end{align}
\eqlab{T}%
\end{subequations}
Here $n\in\Fk$ is a shorthand for $\ket{u\bnk}\in\Fk$,
$\partial_\al = \nabla_{k_\al}$, $\r_{ln\k}$ are the off-diagonal
matrix elements of the position operator, given by
\beq
\r_{ln\k} =
i\ip{u\blk}{{\boldsymbol\nabla}_\k u\bnk} =
\displaystyle
\frac{i\me{u\blk}{\left( {\boldsymbol\nabla}_\k H_\k \right)}{u\bnk}}
{\enk - \elk}
\quad
(\elk\not=\enk)
\,,
\eqlab{r}
\eeq
and
\beq
\begin{aligned}
P_\k &= \sum_{n\in\Fk}\,\ket{u\bnk}\bra{u\bnk}\,,
\\
\one - P_\k
&= \sum_{l\notin\Fk}\,\ket{u\blk}\bra{u\blk}
\end{aligned}
\eqlab{P-Q}
\eeq
are projection operators onto $\Fk$ and onto the complement space.
Throughout the paper, objects such as $T_p$ and $P$ are associated
with $\Fk$, but for brevity we label them with just $\k$.

When $\Fk$ contains the filled states at zero temperature, the
notation
\beq
\left< A \right>
\equiv
\frac{1}{n_{\rm e}}
\int_\k A(\k)
\eqlab{avg}
\eeq
indicates an average over all filled states across the Brillouin zone
(BZ), with $n_{\rm e}$ the electron density and
$\int_\k \equiv \int_{\rm BZ} d^d k/(2\pi)^d$ in $d$ dimensions; we
shall refer to $A(\k)$ and $\langle A \rangle$ as local and global
quantities in reciprocal space, respectively.  We will find it
convenient to sometimes work with BZ-integrated (rather than
BZ-averaged) quantities, denoted with a calligraphic symbol,
\beq
{\mathcal A}
\equiv
\int_\k A(\k)
=
n_{\rm e} \left< A \right>\,.
\eqlab{int}
\eeq

The tensors $T_p(\k)$ are Hermitian, with real-symmetric and
imaginary-antisymmetric parts,
\beq
T_p(\k) = T'_p(\k) +i T''_p(\k)\,.
\eeq
$T_0(\k)$ is an intrinsic property of the Bloch manifold $\Fk$, while
$T_{p\not=0}(\k)$ also depends on the Hamiltonian.  $T_{p\geq 0}(\k)$
remains invariant under multiband unitary transformations within
$\Fk$, and can be recast in a manifestly gauge-invariant form, as
shown in \aref{T-tr-der},
\begin{align}
T_{p \geq 0}^\ab(\k) =
\sum_{l=0}^p\, (-1)^l
\begin{pmatrix}
p\\
l
\end{pmatrix}
\Tr
\left[
P_\k
\left( \partial_\alpha P_\k \right)
H_\k^{p-l}
\left( \partial_\beta P_\k \right)
H_\k^l
\right]\,.
\eqlab{T-tr}
\end{align}
The gauge invariance of this expression follows from the fact that its
ingredients, the Hamiltonian $H_\k$ and the projector $P_\k$, are
obviously unaffected by any unitary mixing among the states
$\ket{u\bnk}$ belonging to $\Fk$.

When $\Fk$ is the ground state, $T_{p\geq 0}(\k)$ is a ground-state
quantity; this is not so for $T_{p< 0}(\k)$, which in fact cannot be
written, in general, in closed form~\cite{traini-epj96}.  The BZ
integral of $T_p(\k)$ satisfies the sum rule
\beq
\frac{2}{\pi}\int_0^\infty d\w\, \w^{p-1}
\sigma_{{\rm er}}^{\rm H}(\w) =
\frac{2 e^2}{\hbar^{p+1}}\int_\k T_p(\k)
\equiv
S_p = S_p' + iS_p''\,,
\eqlab{sum-rule}
\eeq
with $\sigma_{{\rm er}}^{\rm H}$ the Hermitian (absorptive) part of
the interband optical conductivity in the electric-dipole
approximation.  \Eq{sum-rule} yields two sum rules for each $p$: one
symmetric and time even ($S_p'$), the other antisymmetric and time odd
($S_p''$).  The sum rules for $p=1,0,-1$ are summarized in
Table~\ref{tab:sum-rules}; those with $p=0,1$ were obtained in
Refs.~\cite{souza-prb00,souza-prb08}, and the derivation for
arbitrary~$p$ proceeds along similar lines~\cite{verma-pnas25,
  shinada-prb25,hong-unpublished}.  The corresponding atomic sum rules
are given in Refs.~\cite{fano-rmp68,bethe-book86,traini-epj96} for
linearly-polarized light; they diverge for the hydrogen atom when
$p\geq 4$~\cite{bethe-book86}.
\begin{table*}
\centering
\begin{tabular}{lcc}
\hline\hline
\addlinespace
$\phantom{-}p$ & $S_p'$ & $S_p''$\\
\addlinespace
\hline
\addlinespace
$\phantom{-}1$ & $e^2 n_{\rm e}\left< m^{-1}_{\rm er}\right>$ &
$(2|e|/\hbar) n_{\rm e} \langle \mm \rangle$\\
\addlinespace
& Oscillator strength & Rotatory strength\\
\\ 
$\phantom{-}0$ & $(2e^2/\hbar) n_{\rm e}\langle g \rangle$ &
$
-(e^2/\hbar) n_{\rm e} \langle \Omega \rangle$ \\
\addlinespace
& Localization length & Anomalous Hall conductivity 
\\
\\
$-1$ & $\epsilon_0 \chi(0)$ & -- \\
\addlinespace
& Electric susceptibility & ?\\
\addlinespace
\hline\hline
\end{tabular}
\caption{Interband optical sum rules for some values of the integer
  $p$ in \eq{sum-rule}.  $-|e|$ is the electron charge, $n_{\rm e}$ is
  the electron density, $m^{-1}_{\rm er}$ is the interband inverse
  effective mass, $\mm $ is the intrinsic orbital magnetic moment, $g$
  is the quantum metric, and $\Omega$ is the Berry curvature.  Angle
  brackets denote averages over the filled states at zero temperature.
  In insulators $\left< m^{-1}_{\rm er} \right>$ is equal to the
  inverse optical mass $\left< m^{-1}_* \right>$, $\left< g \right>$
  becomes the many-electron localization tensor $\ell^2$, $\chi(0)$
  becomes the clamped-ion electric susceptibility, and the anomalous
  Hall conductivity becomes quantized. The question mark indicates
  that the meaning of $S''_{-1}$ remains unclear. The time-odd
  quantities $S''_p$ vanish in nonmagnetic crystals.}
\label{tab:sum-rules}
\end{table*}

In the next section, we analyze in more detail the tensors $T_0(\k)$
and $T_1(\k)$, together with another composite tensor $R(\k)$.

\section{Survey of composite property tensors}
\seclab{survey}

\subsection{Metric-curvature tensor $T_0(\k)$}
\seclab{T0}

Setting $p=0$ in \eqs{T}{T-tr} gives the metric-curvature
tensor~\cite{berry89,marzari-prb97}
\beq
T^\ab_0(\k) = \sum_{n\in\Fk}\,
\me{\partial_\al u\bnk}{\one - P_\k}{\partial_\be u\bnk} =
\Tr
\left[
P_\k
\left( \partial_\alpha P_\k \right)
\left( \partial_\beta P_\k \right)
\right]
\,,
\eqlab{T0-def}
\eeq
whose real and imaginary parts contain the net quantum
metric~\cite{provost80} and Berry curvature~\cite{berry84} of the
$\Fk$ manifold,
\beq
T_0(\k) = g(\k) - \frac{i}{2}\Omega(\k)\,.
\eqlab{T0}
\eeq
Inserting \eq{P-Q} for $P_\k$ in the first line of \eq{T0-def} yields
\beq
\Omega_\ab(\k) = -2\Im\sum_{n\in\Fk}\,
\ip{\partial_\al u\bnk}{\partial_\be u\bnk}
\equiv
\epsilon_{\alpha\beta\gamma}\Omega_\gamma(\k)
\eqlab{curv}
\eeq
for the curvature (converted to vector form on the right-hand side),
and
\beq
g_\ab(\k) =
\Re\sum_{n\in\Fk}\, \ip{\partial_\al u\bnk}{\partial_\be u\bnk} -
\sum_{ln\in\Fk}
\ip{\partial_\al u\bnk}{u\blk} \ip{u\blk}{\partial_\be u\bnk}
\eqlab{metric}
\eeq
for the metric. Note that the net curvature of a group of states is
the sum of the curvatures of the individual states, while the metric
is not band additive. More generally, the band-additive tensors are
$T''_{2p}(\k)$ and $T'_{2p+1}(\k)$.

The Berry curvature integrated over all filled states gives the
intrinsic anomalous Hall
conductivity~\cite{xiao-rmp10,vanderbilt-book18},
\beq
\sigma^{\rm A}_{{\rm er},\ab}(0) =
-\frac{e^2}{h} 2\pi n_{\rm e} \left< \Omega \right>_\ab
\,;
\eqlab{ahc}
\eeq
the notation on the left-hand side indicates that the intrinsic
anomalous Hall conductivity is the $\w \rightarrow 0$ limit of the
interband part of the antisymmetric optical conductivity. In 2D the
curvature becomes a scalar, $\Omega=\Omega_{xy}=\Omega_z$, and
\eq{ahc} is conveniently written as
\beq
\sigma^{\rm A}_{{\rm er},xy}(0) = -C \frac{e^2}{h}\,.
\eqlab{ahc-2D}
\eeq
Here $C$ is the dimensionless number
\beq
C = 2\pi n_{\rm e} \langle \Omega \rangle\,,
\eqlab{C}
\eeq
which in insulators becomes the integer Chern
number~\cite{vanderbilt-book18}. The corresponding expression in 3D is
\beq
\sigma^{\rm A}_{{\rm er},\ab}(0) = -\epsilon_{\alpha\beta\gamma}
\frac{K_\gamma}{2\pi} \frac{e^2}{h}\,,
\eqlab{ahc-3D}
\eeq
where
\beq
\frac{{\bf K}}{2\pi} =
2\pi n_{\rm e} \langle {\boldsymbol\Omega} \rangle\,.
\eqlab{K}
\eeq
In insulators, ${\bf K}$ becomes a reciprocal lattice vector known as
the Chern vector~\cite{vanderbilt-book18}.\footnote{For simplicity, we
  will continue referring to $C$ as the Chern number and to ${\bf K}$
  as the Chern vector in 2D and 3D metallic systems, where those
  quantities are not quantized.}  In \eq{C} $n_{\rm e}$ is the areal
density of electrons, and in \eq{K} it is the volume density.

In insulators, the average quantum metric of the filled states defines
the many-electron localization tensor~\cite{souza-prb00,resta-epjb11}
\beq
\ell^2_\ab = \langle g \rangle_\ab
\quad
\text{(insulators)}
\,.
\eqlab{loc}
\eeq
Its eigenvalues are the squared electron localization lengths along
the principal axes, and the trace is the gauge-invariant part of the
average spread of the valence Wannier functions~\cite{marzari-prb97}.

\subsection{Mass-moment tensor $T_1(\k)$}
\seclab{T1}

Let us split the mass-moment tensor into real and imaginary parts
as~\cite{martins-scipost25}
\beq
T_1(\k) =
\frac{\hbar^2}{2}m^{-1}_{{\rm er}}(\k) +
\frac{i\hbar}{|e|}\mm(\k)\,.
\eqlab{T1}
\eeq
\Eq{T-geom} gives for the imaginary part
\beq
\begin{aligned}
\mm_\ab(\k) &= \frac{|e|}{\hbar} \sum_{n\in \Fk}
\Im\, \me{\partial_\al u\bnk}{H_\k - \enk}{\partial_\be u\bnk}\\
&+
\frac{|e|}{\hbar} \sum_{ln\in \Fk}
\left( \enk - \elk \right) \Im\,
\ip{\partial_\al u\bnk}{u\blk}\ip{u\blk}{\partial_\be u\bnk}
\\
&\equiv \epsilon_{\alpha\beta\gamma}\mm_\gamma(\k)\,.
\eqlab{orb}
\end{aligned}
\eeq
For a degenerate (or single-state) subspace, the nonadditive second
term vanishes and the first reduces to the intrinsic orbital magnetic
moment~\cite{xiao-rmp10}.  We shall refer to $\mm(\k)$ as an orbital
moment even when $\Fk$ is nondegenerate.  Its ground-state average
$\langle \mm \rangle$ equals $-|e|/2$ times the center-of-mass
circulation per electron~\cite{kunes-prb00,souza-prb08}, which is
related to but different from the ground-state orbital magnetization;
using \eq{T-tr}, $n_{\rm e} \langle \mm \rangle$ becomes the
difference~\cite{souza-prb08} between two gauge-invariant quantities
that add up to the bulk orbital magnetization~\cite{ceresoli-prb06}.

The real part of $T_1(\k)$ gives the interband contribution to the
inverse optical effective mass tensor $m^{-1}_*(\k)$ entering the
oscillator-strength sum rule (\aref{f-sum-rule}).  That tensor has an
additional intraband part --~the transport effective mass~-- so that
\beq
m^{-1}_*(\k) =
m^{-1}_{\rm er}(\k) +
m^{-1}_{\rm ra}(\k)
\eqlab{masses}
\eeq
in total. The three inverse-mass tensors read
\begin{align}
m^{-1}_{*,\ab}(\k) &=
\frac{1}{\hbar^2}
\sum_{n\in \Fk}\,\me{u\bnk}{\left( \partial^2_\ab H_\k \right)}{u\bnk}\,,
\eqlab{mass-opt}
\\
m^{-1}_{{\rm er},\ab}(\k) &=
\frac{2}{\hbar^2} \sum_{n\in\Fk}
\Re\,\me{\partial_\al u\bnk}{H_\k-\enk}{\partial_\be u\bnk}\,,
\eqlab{mass-er}
\\
m^{-1}_{{\rm ra},\ab}(\k) &=
\frac{1}{\hbar^2}
\sum_{n\in \Fk}\,\partial^2_{\ab} \enk\,,
\eqlab{mass-ra}
\end{align}
and the proof that the last two add up to the first is given in
\aref{f-sum-rule}. The expression in \eq{mass-er} for the interband
inverse mass corresponds to the real part of \eq{T-geom} with $p=1$.

For the microscopic Hamiltonian~\eqref{eq:H-loc}, \eq{mass-opt}
becomes
\beq
m^{-1}_{*,\ab}(\k) =
\frac{N_\k}{m_{\rm e}}\delta_{\al\be}
\eqlab{mass-free}
\eeq
with $N_\k$ the number of states in $\Fk$ and $m_{\rm e}$ the
free-electron mass, so that
$\left< m^{-1}_* \right> = m^{-1}_{\rm e}$. The more general form in
\eq{mass-opt} is needed when dealing with nonlocal
pseudopotentials~\cite{martins-scipost25} or low-energy effective
Hamiltonians~\cite{onishi-prx24}.  When $H_\k$ is linear in $\k$ --~as
in $\k\cdot{\bf p}$ models of graphene and of Weyl semimetals~-- the
inverse optical mass vanishes identically, so that
\beq
m^{-1}_{\rm ra}(\k) = -m^{-1}_{\rm er}(\k)
\quad
\text{(linear $H_\k$)\,.}
\eqlab{linear-H}
\eeq

The integral over filled states of the inverse transport mass defines
the Drude weight~\cite{resta-arxiv17},
\beq
D_\ab = \pi e^2 n_{\rm e}\langle m^{-1}_{\rm ra} \rangle_\ab =
\pi e^2 n_{\rm e}\langle m^{-1}_* \rangle_\ab
- \pi e^2 n_{\rm e}\langle m^{-1}_{\rm er} \rangle_\ab\,.
\eqlab{drude}
\eeq
In insulators, where $D_\ab=0$, the two terms on the right-hand side
are equal and opposite.

\subsection{Mass-magnetization tensor $R(\k)$}
\seclab{mass-magnetization}

Motivated by a recent work~\cite{shinada-prb25}, we also consider the
mass-magnetization tensor introduced by
Resta~\cite{resta-arxiv17,resta-jpcm18},
\beq
R_\ab(\k) =
\Tr
\left[
\left| H_\k - \mu \right|
\left( \partial_\alpha P_\k \right)
\left( \partial_\beta P_\k \right)
\right]\,.
\eqlab{R-def}
\eeq
Here $P_\k$ spans all states below the Fermi level $\mu$, so that
\beq
\left| H_\k - \mu \right|
\equiv
\left( H_\k - \mu \right)
\left({\mathbbm 1} - 2P_\k \right)
\eeq
satisfies
\beq
\me{v}{\left| H_\k - \mu \right|}{v} \geq 0
\quad
\text{for all $\ket{v}$}\,.
\eqlab{norm}
\eeq

The tensor $R(\k)$ resembles the metric-curvature
tensor~\eqref{eq:T0-def}. It has the same real part as the mass-moment
tensor~\eqref{eq:T1}, but the imaginary part is different,
\beq
R(\k) =
\frac{\hbar^2}{2}m^{-1}_{{\rm er}}(\k) -
\frac{i\hbar}{|e|}M(\k)\,.
\eqlab{R}
\eeq
Upon integrating over the BZ, the real part gives minus the interband
Drude weight, while the imaginary part
\beq
M_\ab(\k) = \frac{|e|}{\hbar}\sum_{n\in \Fk}
\Im\, \me{\partial_\al u\bnk}{H_\k + \enk - 2\mu}{\partial_\be u\bnk}
\equiv
\epsilon_{\alpha\beta\gamma}M_\gamma(\k)
\eqlab{M-k}
\eeq
[compare with $\mm_\ab(\k)$ in \eq{orb}] gives the ground-state
orbital magnetization,
\beq
{\bf M}_{\rm orb} = n_{\rm e}\langle {\bf M} \rangle\,.
\eqlab{Morb}
\eeq

\section{Positive semidefiniteness}
\seclab{posdef}

If a system is in the ground state, the power absorbed from the
electromagnetic field is necessarily nonnegative,
\beq
\int dt\, {\bf j}(t) \cdot {\bf E}(t) =
\int \frac{d\w}{2\pi}\, {\bf E}^\dagger(\w)
\cdot
\sigma^{\rm H}(\w)
\cdot
{\bf E}(\w) \geq 0\,.
\eqlab{power}
\eeq
For this condition to hold at all frequencies and polarizations of
light, the tensor $\sigma^{\rm H}(\w)$ must be positive
semidefinite. In insulators that tensor is purely interband, while in
metals it has interband and intraband parts that are separately
positive semidefinite.

The spectral moments $S_p$ defined by the interband sum rule in
\eq{sum-rule} inherit the positive semidefiniteness of
$\sigma^{\rm H}_{\rm er}(\w)$.  In fact, the contribution to
$\sigma^{\rm H}_{\rm er}(\w)$ from each $\k$ point already has that
property; and since the sum rule holds separately for each $\k$, the
tensor $T_p(\k)$ is positive semidefinite when $\Fk$ is the ground
state. This property can be seen directly from \eq{pos-def} below.

The subspace of interest is not always the ground state. An example is
a group of bands near the Fermi level, such as the $\pi$ bands in
graphene.  To discuss such cases, consider a generic
positive-semidefinite Hermitian matrix $A$ representing a Cartesian
tensor in~$d$ dimensions,
\beq
{\bf x}^\dagger \cdot A\cdot {\bf x} \geq 0
\text{ for all } {\bf x} \in {\mathbbm C}^d\,.
\eqlab{pos-def-def}
\eeq
For $A = T_p(\k)$, \eq{T-sos1} gives
\begin{align}
{\bf x}^\dagger \cdot T_p(\k) \cdot {\bf x} &=
\sum_{\substack{n\in\Fk\\{l\notin F_\k}}}
\left| {\bf x} \cdot {\bf r}_{ln\k} \right|^2
\left( \elk - \enk \right)^p\,.
\eqlab{pos-def}
\end{align}
When $p$ is even, this expression is non-negative for all ${\bf x}$,
so that both $T_p(\k)$ and its BZ integral $S_p$ are positive
semidefinite; in particular, for $p=0$ one recovers the positive
semidefiniteness of the metric-curvature
tensor~\cite{roy-prb14,peotta-natcomms15,ozawa-prb21}.  When $p$ is
odd, as in the case of the mass-moment tensor, $T_p(\k)$ and $S_p$ are
not guaranteed to be positive semidefinite if there are lower-energy
states at $\k$ outside $\Fk$. When that happens, we say that $\Fk$ is
a high-lying manifold; otherwise we call it a low-lying manifold.  The
mass-magnetization tensor, which by construction describes the
low-lying manifold of filled states, is positive semidefinite by
virtue of \eq{norm}~\cite{shinada-prb25}.

When a Hermitian matrix $A$ is positive semidefinite, its real part
has the same property,
\beq
{\bf x}^{\rm T} \cdot A' \cdot {\bf x} \geq 0
\text{ for all } {\bf x} \in {\mathbbm R}^d\,,
\eqlab{pos-def-re}
\eeq
as can be seen by writing
\beq
\begin{aligned}
{\bf x}^{\rm T} \cdot A' \cdot {\bf x} &=
{\bf x}^{\rm T} \cdot
\frac{1}{2}\left( A + A^{\rm T} \right) \cdot {\bf x}\\
&=
\frac{1}{2} \left( {\bf x}^{\rm T} \cdot A \cdot {\bf x} \right) +
\frac{1}{2} \left( {\bf x}^{\rm T} \cdot A \cdot {\bf x} \right)^{\rm T}\\
&= {\bf x}^{\rm T} \cdot A \cdot {\bf x}\,.
\end{aligned}
\eeq
When $A$ is the metric-curvature tensor, \eq{pos-def-re} states that
the quantum metric of a generic manifold of states is positive
semidefinite~\cite{provost80}. When $A$ is either the mass-moment or
the mass-magnetization tensor, it states that the interband inverse
mass traced over a low-lying manifold is positive semidefinite.

The positive semidefiniteness --~under appropriate conditions~-- of
the various composite tensors introduced in the previous section
imposes certain relations among the associated electronic properties.
We shall consider matrix-invariant inequalities in two and three
dimensions (\srefs{roy}{roy3d}), followed by gap inequalities and
Cauchy-Schwarz inequalities and their combination with
matrix-invariant inequalities (\sref{spectral}).

\section{2D matrix-invariant inequalities}
\seclab{roy}

\subsection{General relations}

A necessary and sufficient condition for a Hermitian matrix $A$ to be
positive semidefinite is that all its principal minors are
nonnegative~\cite{prussing86}.  When $A$ represents a 2D Cartesian
tensor, that condition reads
\beq
\begin{aligned}
\det A = A_{xx}A_{yy} - \left| A_{xy} \right|^2 &\geq 0\,,\\
A_{xx} \geq 0\,,\;
A_{yy} &\geq 0\,,
\eqlab{sylvester}
\end{aligned}
\eeq
and it imposes certain relations among the scalar invariants of $A$.
When $A$ is the metric-curvature tensor, those relations take the form
of inequalities involving the quantum metric and Berry curvature
tensors, first discussed by Roy~\cite{roy-prb14}. The basic arguments
are given below in a form that will facilitate their extension to 3D.

Writing $A=A'+iA''$ ($A'$ is also positive semidefinite), the first
condition in \eq{sylvester} becomes
$\left( A''_{xy} \right)^2 \leq \det\, A'$, and invoking the other two
conditions one finds
\beq
\left| A''_{xy} \right|
\leq
\sqrt{\det\, A'}
\leq
\sqrt{A'_{xx}\, A'_{yy}}
\leq
\frac{1}{2} \left( A'_{xx} + A'_{yy} \right)
\,,
\eqlab{chained}
\eeq
where $A'_{\al\al} = A_{\al\al}$. The middle inequality saturates when
$x$ and $y$ are the principal axes of $A'$. The right inequality,
which saturates when $A'_{xx}=A'_{yy}$, states that the geometric mean
of two nonnegative numbers cannot exceed their arithmetic mean.

\Eq{chained} depends on the choice of coordinates. To obtain invariant
relations, note that a $2\times 2$ Hermitian matrix $A$ has three
scalar invariants:\footnote{The invariants of a Hermitian matrix
  $A = A' + iA''$ are the same as those of the nonsymmetric real
  matrix $A'+A''$, which are discussed in
  Ref.~\cite{enwiki:1269822116}.}  the norm of the vector ${\bf A}''$
with components $\left( 0, 0, A''_{xy} \right)$, and the two principal
invariants of the real part.  The principal
invariants~\cite{enwiki:1269822116,holzapfel-book00} are the
coefficients of the characteristic polynomial
$\det\left(A'-\lambda {\mathbbm 1}\right)$, which in 2D takes the form
$\lambda^2 - {\mathcal I}_1 \lambda + {\mathcal I}_2$.  Thus,
\beq
\begin{aligned}
{\mathcal I}_1 &= \tr\, A' = \lambda_1 + \lambda_2\,,\\
{\mathcal I}_2 &= \det\, A' = \lambda_1\lambda_2\,.
\eqlab{I1-I2}
\end{aligned}
\eeq
One can now identify in \eq{chained} the invariant relations
\beq
\left| {\bf A}'' \right|
\leq
\sqrt{\det\, A'}
\leq
\frac{1}{2}\tr\, A'\,,
\eqlab{ineqs-2D}
\eeq
where the right inequality relates the geometric and arithmetic means
of the non-negative eigenvalues of $A'$.  Note that the outer
inequality follows directly from the positive semidefiniteness
condition~\eqref{eq:pos-def-def} as~\cite{roy-prb14}
\beq
\begin{bmatrix}
1 & \mp i
\end{bmatrix}
\cdot A \cdot
\begin{bmatrix}
1\\
\pm i
\end{bmatrix}
\geq 0\,.
\eqlab{circ-pol}
\eeq

\subsection{Bound saturation and optical absorption}
\seclab{saturation}

When $A$ is the spectral moment $S_p$ defined by \eq{sum-rule} for the
ground-state configuration, condition~\eqref{eq:circ-pol} can be
linked to the non-negativity of interband absorption from
circularly-polarized light~\cite{onishi-prx24}. This line of reasoning
allows to discuss the saturation of the sum-rule inequalities
\beq
\left| {\bf S}_p'' \right|
\leq
\sqrt{\det S_p'}
\leq
\frac{1}{2}\tr\, S_p'
\eqlab{ineqs-Sp}
\eeq
from the perspective of optical absorption.

The positive semidefiniteness condition on $S_p$, from which
\eq{ineqs-Sp} follows, reads
\beq
{\boldsymbol \epsilon}^\dagger
\cdot
S_p
\cdot
{\boldsymbol \epsilon} \geq 0
\Leftrightarrow
\int_0^\infty d\w\, \w^{p-1}
{\boldsymbol \epsilon}^\dagger
\cdot
\sigma^{\rm H}_{\rm er}(\w)
\cdot
{\boldsymbol \epsilon}
\geq 0\,,
\eqlab{Sp-posdef-int}
\eeq
where ${\boldsymbol\epsilon}$ is the in-plane polarization vector of
the optical field: see \eq{power} for the absorbed power. According to
the previous subsection, the left inequality in \eq{ineqs-Sp} amounts
to the condition $\det S_p \geq 0$, and therefore it saturates if and
only if $S_p$ is singular. In view of \eq{Sp-posdef-int}, $S_p$ being
singular means that there is at least one polarization vector
${\boldsymbol \epsilon}$ --~the eigenvector of $S_p$ with null
eigenvalue~-- for which the absorbed power vanishes at all
frequencies.  If that polarization is circular, the outer inequality
in \eq{ineqs-Sp} saturates as well, see \eq{circ-pol}, which means
that the system exhibits 100\% magnetic circular
dichroism~\cite{onishi-prx24}. For that to happen, the tensor $S'_p$
must be isotropic, so that the right inequality saturates.

Two systems where the inequalities in \eq{ineqs-Sp} are saturated or
nearly so will be studied in the last part of this section.  In
\sref{spectral}, we will encounter other types of sum-rule
inequalities, and discuss the conditions on interband absorption
leading to their saturation.

\subsection{Local and global inequalities}
\seclab{loc-glob-2D}

When the matrix $A$ depends on $\k$, the inequalities in \eq{ineqs-2D}
hold locally in $k$ space,
\beq
\left| {\bf A''(\k)} \right|
\leq
\sqrt{\det\, A'(\k)}
\leq
\frac{1}{2}\tr\, A'(\k)\,.
\eqlab{local-2D}
\eeq
For $A(\k) = T_p(\k)$, the left inequality saturates in two-band
models; this result is demonstrated in Ref.~\cite{ozawa-prb21}, and an
alternative proof is given in \aref{saturated}.

The global matrix ${\mathcal A}$ defined by \eq{int} inherits the
positive semidefiniteness of the local $A(\k)$, leading to
\beq
\left| \boldsymbol{\mathcal A}'' \right|
\leq
\sqrt{\det {\mathcal A}'}
\leq
\frac{1}{2}\tr\,{\mathcal A'}\,,
\eqlab{global-2D-a}
\eeq
where
\beq
\begin{aligned}
\boldsymbol{\mathcal A}''
&=
\int_\k {\bf A}''(\k)\,,
\\
\tr\, {\mathcal A}'
&=
\int_\k \tr\, A'(\k)\,,
\\
\det {\mathcal A}'
&=
\det \int_\k A'(\k)
\geq
\int_\k  \det A'(\k)\,.
\end{aligned}
\eqlab{BZ-defs}
\eeq
The inequality follows from $\det (A+B) \geq \det\, A + \det\, B$
for $A$ and $B$ positive semidefinite.

Combining \eq{local-2D} with the triangle inequality
$\left| \boldsymbol{\mathcal A}'' \right| \leq \int_\k \left|
A''_z(\k) \right|$ yields
\beq
\left| \boldsymbol{\mathcal A}'' \right|
\leq
\int_\k \sqrt{\det\, A'(\k)}
\leq
\frac{1}{2} \tr\, {\mathcal A}'\,,
\eqlab{global-2D-b}
\eeq
which differ from \eq{global-2D-a} by the middle term. The Minkowski
determinant inequality~\cite{mirsky-book55} applied to $2\times2$
positive-semidefinite matrices,
\beq
\sqrt{\det\, A}+
\sqrt{\det\, B}
\leq
\sqrt{\det\,(A+B)}\,,
\eqlab{minkowski}
\eeq
implies
\beq
\int_\k \sqrt{\det\, A'(\k)}
\leq
\sqrt{\det\, {\mathcal A}'}
\,.
\eqlab{I2-minkowski}
\eeq
This allows to combine \eqs{global-2D-a}{global-2D-b} into a single
set of global relations,
\beq
\left| \boldsymbol{\mathcal A}'' \right|
\leq
\int_\k \sqrt{\det\, A'(\k)}
\leq
\sqrt{\det\, {\mathcal A}'}
\leq
\frac{1}{2} \tr\, {\mathcal A}'\,.
\eqlab{global-2D}
\eeq
For $A(\k) = T_p(\k)$, in which case ${\mathcal A}\propto S_p$ and
\eq{global-2D} becomes \eq{ineqs-Sp}, the first inequality saturates
in two-band models if $A_z''(\k)$ does not change sign across the
BZ~\cite{ozawa-prb21}. This follows because under those conditions
both the triangle inequality and the left inequality in \eq{local-2D}
saturate.

Below, we apply the local and global inequalities~\eqref{eq:local-2D}
and~\eqref{eq:global-2D} to the metric-curvature, mass-moment, and
mass-magnetization tensors.

\subsection{Metric-curvature inequalities}
\seclab{metric-curv-2D}

When $A(\k)$ is the metric-curvature tensor~\eqref{eq:T0}, the local
inequalities~\eqref{eq:local-2D} become the Roy inequalities between
the magnitude of the 2D Berry curvature and the determinant and trace
of the 2D quantum
metric~\cite{roy-prb14,peotta-natcomms15,ozawa-prb21},
\beq
\left| \Omega(\k) \right|
\leq
2\sqrt{\det g(\k)}
\leq
\tr\, g(\k)\,.
\eqlab{roy}
\eeq
These relations hold for both low- and high-lying manifolds.  As
already mentioned, the left inequality saturates in two-band
models~\cite{ozawa-prb21}; deviations from saturation due to couplings
to distant bands were studied in Ref.~\cite{martins-scipost25} using
an {\it ab initio} model of gapped graphene where inversion symmetry
is broken by a staggered sublattice potential.

Turning now to global quantities, when $A(\k)$ is the metric-curvature
tensor we have
\beq
\begin{aligned}
{\mathcal A}_z'' &=
-\frac{1}{2} n_{\rm e} \langle \Omega \rangle\,,
\\
\tr\, {\mathcal A}' &=
n_{\rm e}\, \tr\, \langle g \rangle\,,
\\
\det {\mathcal A}' &=
n_{\rm e}^2 \det\, \langle g \rangle\,.
\end{aligned}
\eeq
Using these expressions, \eq{global-2D} becomes
\beq
\left| C \right|
\leq
4\pi n_{\rm e} \left< \sqrt{\det g} \right>
\leq
4\pi n_{\rm e} \sqrt{\det\, \langle g \rangle}
\leq
2\pi n_{\rm e}\, \tr\, \langle g \rangle\,.
\eqlab{ineqs-p0}
\eeq
This set of inequalities is given in Ref.~\cite{ozawa-prb21} for
insulators. In that case $C$ is the integer Chern number,
$\langle g \rangle$ is the localization tensor~$\ell^2$, and the
middle inequality relate the quantum volumes (volumes measured with
the quantum metric) in $k$ space and in twist-angle
space~\cite{ozawa-prb21}.

\subsection{Mass-moment inequalities}
\seclab{bounds-m-2D}

When $A(\k)$ is the mass-moment tensor~\eqref{eq:T1}, \eq{local-2D}
gives upper bounds on the magnitude of the orbital magnetic moment of
a low-lying Bloch manifold,
\beq
\left| \mm(\k) \right|
\leq
\sqrt{\det\, \mu_{\rm B}^{\rm er}(\k)}
\leq
\frac{1}{2}\tr\,\mu_{\rm B}^{\rm er}(\k)\,,
\eqlab{ineq-m}
\eeq
where
\beq
\mu_{\rm B}^{\rm er}(\k) = \frac{|e|\hbar}{2} m^{-1}_{\rm er}(\k)
\eqlab{mu-er}
\eeq
is an effective magneton tensor defined in terms of the interband
effective mass tensor~\eqref{eq:mass-er}, which is positive
semidefinite for a low-lying manifold (\sref{posdef}).

The replacement of $m^{-1}_{\rm er}(\k)$ in \eq{mu-er} with
$m^{-1}_{\rm ra}(\k)$ and $m^{-1}_*(\k)$ defines intraband and optical
effective magneton tensors $\mu_{\rm B}^{\rm ra}(\k)$ and
$\mu_{\rm B}^*(\k)$.  In two-band models the left inequality in
\eq{ineq-m} saturates (\aref{saturated}), and if the Hamiltonian
$H_\k$ is linear then
$\mu_{\rm B}^{\rm er}(\k) = -\mu_{\rm B}^{\rm ra}(\k)$, so that
\beq
\left| \mm(\k) \right| = \sqrt{\det\, \mu_{\rm B}^{\rm ra}(\k)}
\,.
\eeq
If moreover the energy dispersion becomes isotropic for
$\k\rightarrow \0$, the right inequality in \eq{ineq-m} saturates as
well, yielding
\beq
\left| \mm(\0) \right| = \left| \mu_{\rm B}^{\rm ra}(\0) \right| =
\frac{|e|\hbar}{2|m_{\rm ra}(\0)|}\,.
\eqlab{m-xiao}
\eeq
This result was obtained in Ref.~\cite{xiao-prl07} for the orbital
moment in a low-energy model of gapped graphene. The present
derivation clarifies that it is a special case of a general bound
relation.

The global relations in \eq{global-2D} give upper bounds on the
magnitude of the orbital moment averaged over all filled states,
\beq
\left| \left< \mm \right> \right|
\leq
\left<
\sqrt{\det\, \mu_{\rm B}^{\rm er}}
\right>
\leq
\sqrt{\det \left< \mu^{\rm er}_{\rm B} \right>}
\leq
\frac{1}{2} \tr \left< \mu^{\rm er}_{\rm B} \right>\,.
\eqlab{ineq-m2}
\eeq
In insulators,
$\left< \mu^{\rm er}_{\rm B} \right>=\left< \mu_{\rm B}^{*} \right>$.
In metals, using $\left< \mu^*_{\rm B} \right>$ in \eq{ineq-m2} leads
to weaker bounds on $\left| \left< \mm \right> \right|$, because the
added amount $\left< \mu^{\rm ra}_{\rm B} \right>$, proportional to
the Drude weight, is positive semidefinite.  For a microscopic
Hamiltonian, $\left< \mu^*_{\rm B} \right>$ becomes the true Bohr
magneton $\mu_{\rm B}=|e|\hbar/2m_{\rm e}$, so that
\beq
\left| \left< \mm \right> \right|
\leq
\mu_{\rm B}
\eqlab{bound-m-micro}
\eeq
holds for both insulators and metals.

\subsection{Mass-magnetization inequalities}
\seclab{bounds-M-2D}

When $A(\k)$ is the mass-magnetization tensor~\eqref{eq:R},
\eq{global-2D} yields a set of upper bounds on the magnitude of the 2D
orbital magnetization,
\beq
\left| M_{\rm orb} \right|
\leq
n_{\rm e}
\left<
\sqrt{\det\, \mu_{\rm B}^{\rm er}}
\right>
\leq
n_{\rm e}
\sqrt{\det \left< \mu^{\rm er}_{\rm B} \right>}
\leq
\frac{1}{2}
n_{\rm e}
\tr \left< \mu^{\rm er}_{\rm B} \right>\,.
\eqlab{bounds-M}
\eeq
The intermediate bound on $\left| M_{\rm orb} \right|$ was recently
obtained in Ref.~\cite{shinada-prb25} along with
\beq
\left| M_{\rm orb} \right|
\leq
n_{\rm e} \mu_{\rm B}
\,,
\eeq
which, by the same argument leading to \eq{bound-m-micro}, holds in
both insulators and metals described by a microscopic Hamiltonian.

Let us split the orbital magnetization as~\cite{souza-prb08}
\beq
M_{\rm orb} = n_{\rm e}\langle \mm \rangle + \Delta M\,.
\eqlab{Morb-split}
\eeq
The first term is the part captured by the optical sum rule $S''_1$ in
Table~\ref{tab:sum-rules}, and $\Delta M$ is the remainder.  As shown
above, the magnitudes of the total $M_{\rm orb}$ and of the optical
part $n_{\rm e}\langle\mm\rangle$ satisfy the same upper bounds; this
is a consequence of the mass-moment and mass-magnetization tensors in
\eqs{T1}{R} having the same real part.

\subsection{Illustrative examples}
\seclab{examples-2D}

\subsubsection{Landau levels}
\seclab{landau-levels-orb}

The spectrum of a 2D free electron gas in a transverse magnetic field
consists of flat Landau levels separated by the cyclotron frequency.
The metric-curvature inequalities~\eqref{eq:roy}
and~\eqref{eq:ineqs-p0} are known to saturate for this
system~\cite{ozawa-prb21}.  More generally, when the $\nu$ lowest
Landau levels are full, the local and global
inequalities~\eqref{eq:local-2D} and \eqref{eq:global-2D} saturate for
$A(\k) = T_p(\k)$ with arbitrary $p$: see \aref{LL}. (The optical
conditions for saturation discussed in \sref{saturation} are fulfilled
as well.) In particular, for $p=1$ \eq{bound-m-micro} gives
$\left| \left< \mm \right> \right| = \mu_{\rm B}$.

The behavior of the magnetization bound~\eqref{eq:bounds-M} is
discussed in Ref.~\cite{shinada-prb25}. The bound obtained in that
work, combined with the saturated orbital-moment bound, reads
\beq
\left| M_{\rm orb}(\mu) \right|
\leq
n_{\rm e} \left| \left< \mm \right> \right|
=
n_{\rm e} \mu_{\rm B}\,.
\eqlab{Morb-bound}
\eeq
The chemical potential $\mu$ is assumed to lie between Landau levels
$\epsilon_{\nu-1}$ and $\epsilon_\nu$, so that the electron density is
$n_{\rm e} = (\nu/h)|eB|$ (ignoring spin). As $\mu$ is scanned across
the gap by varying the occupation of the edge states, the sum-rule
part of the orbital magnetization stays constant with magnitude
$n_{\rm e} \mu_{\rm B}$, while the total orbital magnetization varies
linearly, crossing zero in the middle of the gap.  The maximum
magnitude $n_{\rm e} \mu_{\rm B}$ of the total orbital magnetization
is reached when the edge states are either fully occupied
($\mu=\epsilon_{\nu}$) or completely empty ($\mu=\epsilon_{\nu-1}$).

As discussed in \sref{posdef}, positive semi-definiteness of the
odd-$p$ tensors $T_p(\k)$ and their BZ integrals is only guaranteed
for low-lying manifolds.  Hence, bound relations involving those
tensors may not hold for high-lying manifolds. As an example with
$p=1$, the orbital-moment bound in \eq{bound-m-micro} is violated upon
filling a single Landau level with $\nu> 0$: see \aref{LL}. Instead,
the metric-curvature relations ($p=0$) are still satisfied.

\subsubsection{Haldane model}
\seclab{haldane-2D}

\begin{figure}
\centering
\includegraphics[width=0.65\columnwidth]{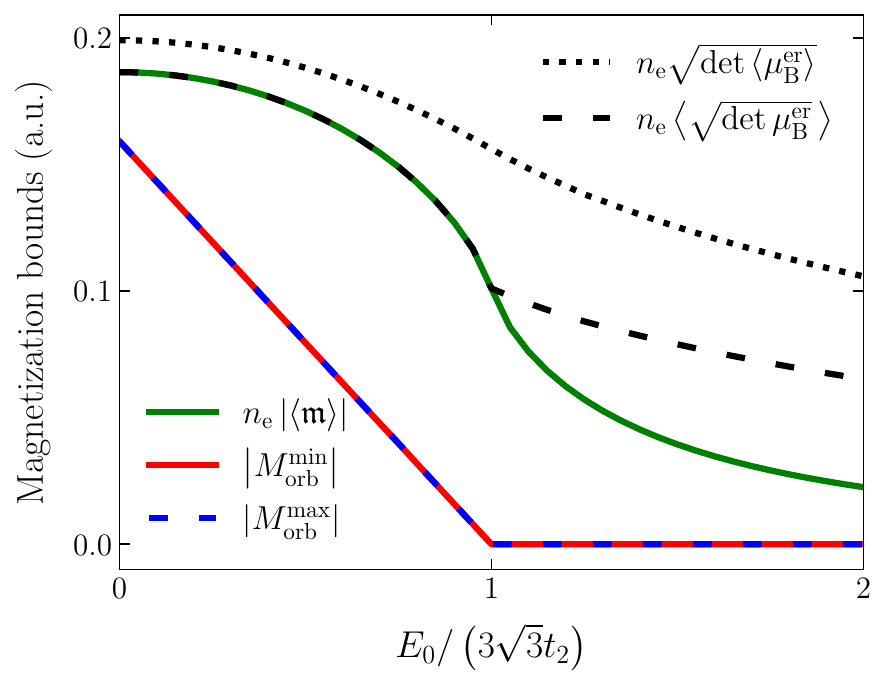}
\caption{Solid and dashed colored lines at the bottom: parametric
  dependence, in the Haldane model with the lower band occupied, of
  the maximum magnitude of the orbital magnetization $M_{\rm orb}$,
  and of the magnitude of its optical part
  $n_{\rm e} \langle \mm \rangle$ (electron density times average
  orbital moment). Dashed and dotted black lines at the top: upper
  bounds on both quantities [\eqs{ineq-m2}{bounds-M}], with
  $\mu^{\rm er}_{\rm B}$ the interband effective magneton of
  \eq{mu-er}. The model parameters are given in the main text.}
\figlab{fig1}
\end{figure}

The Haldane model~\cite{haldane-prl88} is a spinless tight-binding
model on a honeycomb lattice, with energies $\pm E_0$ on the two
inequivalent sites, real first-neighbor hoppings~$t_1$, and complex
second-neighbor hoppings $t_2e^{\pm i\varphi}$.  We adopt Hartree
atomic units (a.u.), $\hbar = m_{\rm e} = |e| = 4\pi\epsilon_0 = 1$.
The lattice constant is set to one Bohr radius $a_0$, $t_1=1$~Ha,
$t_2=1/\sqrt{3}$~Ha, $\varphi=\pi/2$, and the dimensionless ratio
$E_0/t_2$ is treated as an adjustable parameter. When
$\left| E_0/\left( 3\sqrt{3} t_2\right) \right| < 1$, the system with
the lower band occupied is a Chern insulator with $|C|=1$; for
$\left| E_0/\left( 3\sqrt{3} t_2\right) \right| > 1$, it becomes a
trivial insulator with $C=0$. At the phase boundary, the energy gap
closes at a corner point in the hexagonal BZ.

The local and global metric-curvature inequalities~\eqref{eq:roy}
and~\eqref{eq:ineqs-p0} were studied for this model in
Refs.~\cite{roy-prb14} and~\cite{ozawa-prb21}, respectively. Here, we
focus on the global mass-moment and mass-magnetization inequalities
\eqref{eq:ineq-m2} and~\eqref{eq:bounds-M}. The numerical results are
plotted in \fref{fig1} versus $ E_0/\left( 3\sqrt{3} t_2\right)$, with
the chemical potential $\mu$ placed in the energy gap.

To understand the behavior of the two $M_{\rm orb}$ curves in
\fref{fig1}, consider the relation~\cite{vanderbilt-book18}
\beq
\frac{d M_{\rm orb}}{d\mu} = \frac{|e|}{h}C\,,
\eqlab{M-mu}
\eeq
which follows from differentiating \eq{M-k} with respect to $\mu$.
The particle-hole symmetry of the Haldane model with $\varphi=\pi/2$
implies
\beq
M_{\rm orb} = \frac{|e|}{h}\mu C
\quad
\text{($\mu$ inside the gap)}
\,,
\eqlab{M-mu-linear}
\eeq
with $\mu$ measured from the middle of the gap. Thus, $M_{\rm orb}$
vanishes in the trivial phase where $C=0$, and in the topological
phase its magnitude increases linearly with that of $\mu$ inside the
gap, just like in Landau levels at integer filling.  Following
Ref.~\cite{shinada-prb25}, we plot its maximum magnitude
$\left| M_{\rm orb}^{\rm min}\right| = \left|M_{\rm orb}^{\rm max}
\right|$, attained when $\mu$ reaches the top of the valence band or
the bottom of the conduction band.

The solid green line in \fref{fig1} gives the magnitude of
$n_{\rm e}\langle \mm \rangle$ --~the sum-rule part of
$M_{\rm orb}$~-- while the dashed and dotted black lines provide upper
bounds on both $n_{\rm e} \left| \langle \mm \rangle \right|$ and
$\left| M_{\rm orb} \right|$. The weaker bound is also plotted in
Ref.~\cite{shinada-prb25}; that line actually represents two separate
bounds --~the geometric and arithmetic means of the principal values
of $n_{\rm e}\langle \mu^{\rm er}_{\rm B} \rangle$~-- which are
numerically the same because the rotational symmetry of the system
renders that tensor isotropic.

In the topological phase, the tighter upper bound in \fref{fig1}
appears to coincide with
$n_{\rm e} \left| \langle \mm \rangle \right|$.  As discussed below
\eq{global-2D}, that coincidence occurs when $\mm(\k)$ does not change
sign across the BZ. This condition is nearly satisfied in the
topological phase of the model: in that phase, $\mm(\k)$ only changes
sign near $\k = {\bf 0}$, where its magnitude is tiny and the
contribution to the BZ integral negligible. (As the same is true for
$\Omega(\k)$, there is also a near saturation of the tighter bound on
$|C|$ in \eq{ineqs-p0}~\cite{ozawa-prb21}.) At $E_0=0$, the saturation
becomes exact.

\subsubsection{Flat-band model}
\seclab{flatband-magnetization}

\begin{figure}
\centering
\includegraphics[width=0.49\columnwidth]{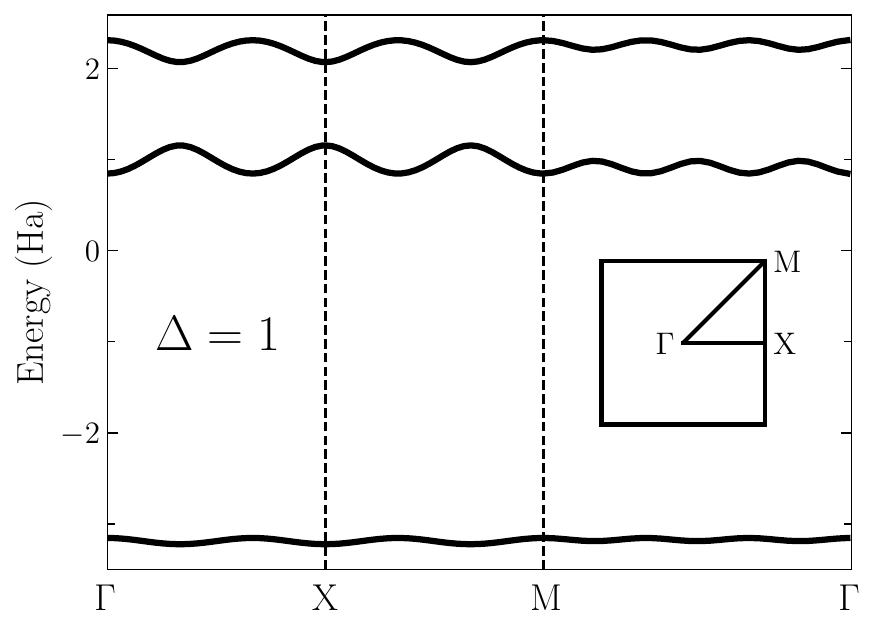}
\includegraphics[width=0.49\columnwidth]{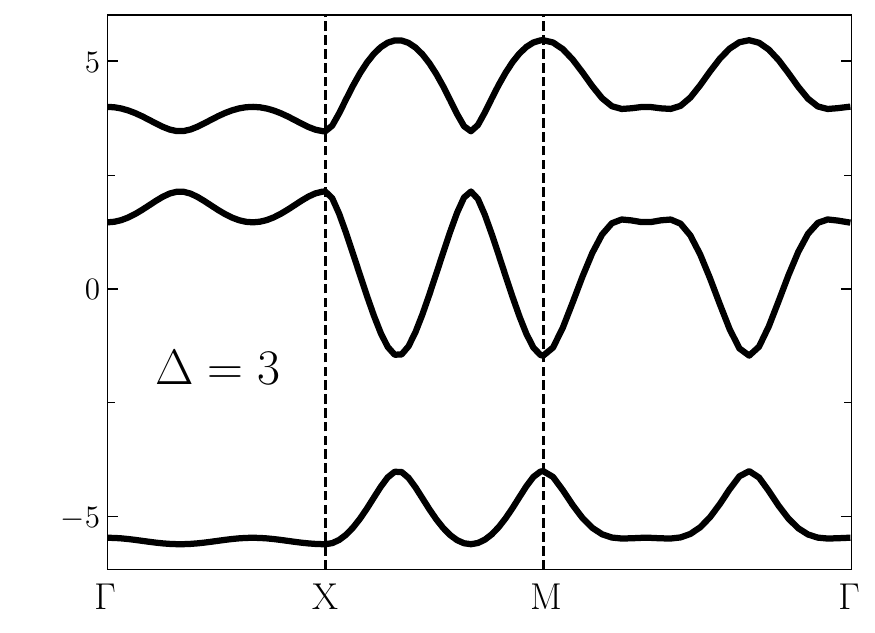}
\caption{Energy bands of a tight-binding model on a square lattice
  with tunable flat bands controlled by a parameter
  $\Delta$~\cite{yang-prb12}. Left: maximally-flat bands for
  $\Delta = 1$. Right: wide bands for $\Delta = 3$.}
\figlab{fig2}
\end{figure}

As discussed in \sref{landau-levels-orb} and in \aref{LL}, the various
bound relations saturate in Landau levels with integer filling and no
band dispersion. The tendency for bound relations to approach
saturation as bands get flatter has been noted in several works for
specific models of Chern insulators. That tendency was investigated
for the global metric-curvature bounds in Ref.~\cite{ozawa-prb21}, and
for the orbital-magnetization bounds in Ref.\cite{shinada-prb25}.

We focus here on a system studied in
Refs.~\cite{ozawa-prb21,shinada-prb25}, a tight-binding model with
three orbitals per site on a square lattice~\cite{yang-prb12}. There
are real and complex first-neighbor hoppings and complex
second-neighbor hoppings, controlled by three parameters, $t_1$, $t_2$
and $\phi$. As in the works cited above, we set $t_1 =1$~Ha,
$t_2=-\Delta/\sqrt{3}$~Ha and $\phi=\pi/3$.  The lower band has Chern
number $C=3$, and band flatness is controlled by the parameter
$\Delta$.  \Fref{fig2} shows the band structure for $\Delta=1$
(maximally-flat bands) and for $\Delta = 3$ (wide bands).  In the
following, the chemical potential is placed in the lower gap.

\Fref{fig3} plots versus $\Delta$ the same quantities shown in
\fref{fig1} for the Haldane model.  At $\Delta=1$, the maximum
magnitude of $M_{\rm orb}$ comes very close to that of its optical
part, and the upper bounds on both quantities nearly saturate.  Away
from $\Delta=1$, all five curves drift apart. Since the model lacks
particle-hole symmetry, $\left| M_{\rm orb} \right|$ differs between
empty and full edge states; this is why the
$\left| M_{\rm orb}^{\rm max}\right|$ and
$\left|M_{\rm orb}^{\rm min} \right|$ curves are different, unlike in
\fref{fig1}.

To understand the convergence of the three magnetization curves at
$\Delta = 1$, recast \eq{Morb} for the orbital magnetization as
\beq
\begin{aligned}
M_{\rm orb} &= \int_\k \mm(\k) +
\frac{|e|}{\hbar}
\int_\k \left[ \mu - \epsilon(\k) \right] \Omega(\k)\\
&\equiv
n_{\rm e} \langle \mm \rangle +
\Delta M
\,;
\end{aligned}
\eqlab{Morb-single}
\eeq
$\epsilon(\k)$ is the energy of the lower (filled) band, and the
second line is the same as \eq{Morb-split}.  Since at $\Delta = 1$ the
lower band is nearly flat, in the second term the quantity in square
brackets can be brought outside the integral, yielding
\beq
\Delta M
\approx
\left( \mu - \epsilon \right) \frac{|e|}{h}C\,.
\eqlab{DeltaM-single-flat}
\eeq
\begin{figure}
\centering
\includegraphics[width=0.65\columnwidth]{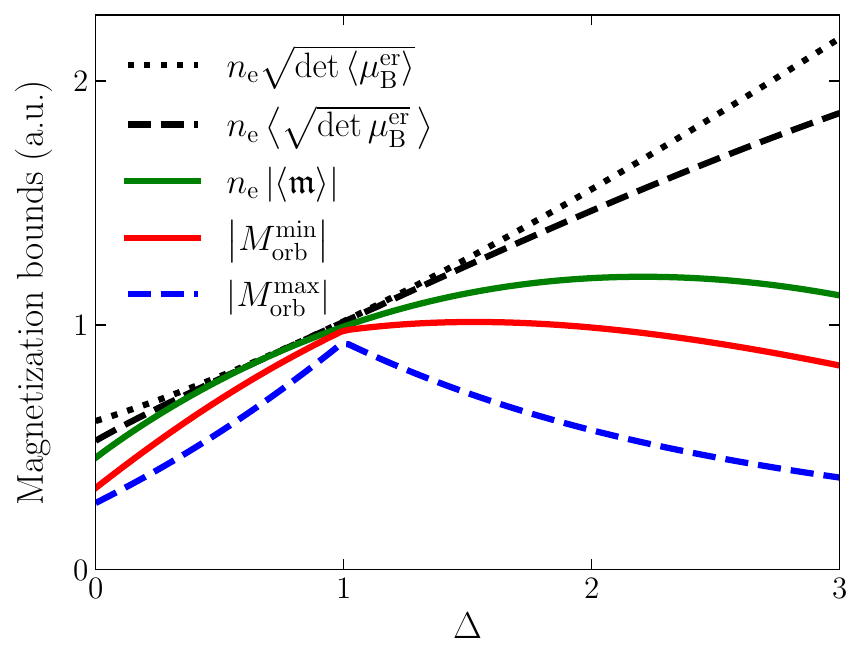}
\caption{Same as \fref{fig1}, but for a three-band tight-binding model
  on a square lattice with tunable flat bands~\cite{yang-prb12}, with
  the lowest band taken as occupied.  As in \fref{fig1}, the dotted
  line represents two separate upper bounds that are degenerate by
  symmetry.}
  \figlab{fig3}
\end{figure}
In addition to band flatness, the convergence of the magnetization
curves in \fref{fig3} depends on another property of the model at
$\Delta = 1$: dipole transitions between the lower and upper bands are
negligible compared to those between the lower and middle bands (this
was checked numerically by evaluating $\left| \r_{ln\k} \right|$ on a
BZ grid). As a result, the orbital moment and Berry curvature of the
lower band satisfy
\beq
\mm(\k) \approx - \frac{|e|}{2\hbar}\Eg(\k)\Omega(\k)\,,
\eqlab{m-curv}
\eeq
as in two-band models. Using \eq{m-curv} in the first term of
\eq{Morb-single} and neglecting the weak dispersion in the direct gap
$\Eg(\k)$, one finds
\beq
n_{\rm e} \langle \mm \rangle
\approx
-\frac{|e|}{h} \frac{\Eg}{2} C \,.
\eqlab{Mopt-single-flat}
\eeq
The sum of \eqs{DeltaM-single-flat}{Mopt-single-flat} gives
\beq
M_{\rm orb} \approx \frac{|e|}{h}
\left[
\mu - \left( \epsilon + \Eg/2 \right)
\right]
C
\,,
\eeq
which has the same form as \eq{M-mu-linear}. Thus,
$\left| M_{\rm orb} \right|$ increases linearly as $\mu$ deviates from
the middle of the lower gap, just like in Landau levels and in the
Haldane model with particle-hole symmetry. Moreover,
\beq
\left| M^{\rm min}_{\rm orb} \right|
\approx
\left| M^{\rm max}_{\rm orb} \right|
\approx
n_{\rm e}
\left| \langle \mm \rangle \right|
\approx
\frac{|eC| \Eg}{2h}
\left( = \frac{|C|\Eg}{4\pi}\text{ in a.u.} \right)\,.
\eqlab{M-approx}
\eeq
Since $|C| = 3$ and $\Eg$ is slightly larger than 4~Ha, the near
coincidence between $\left| M^{\rm min}_{\rm orb} \right|$,
$\left| M^{\rm max}_{\rm orb} \right|$ and
$n_{\rm e} \left| \langle \mm \rangle \right|$ at $\Delta = 1$ occurs
at a magnetization value close to $1$~a.u., as seen in \fref{fig3}.

Consider now the tighter upper bound in \fref{fig3} (dashed black
line) in relation to $n_{\rm e} \left| \langle \mm \rangle
\right|$. Their near coincidence at $\Delta = 1$ results from
$\mm(\k)$ having a constant sign across the BZ (this was checked
numerically), see discussion below \eq{global-2D}. Even though that
discussion was for two-band models, the conclusion holds to a good
approximation in this three-band model at $\Delta = 1$, due to the
negligible electric-dipole coupling between the lower and upper bands.

\begin{figure}
\centering
\includegraphics[width=0.49\columnwidth]{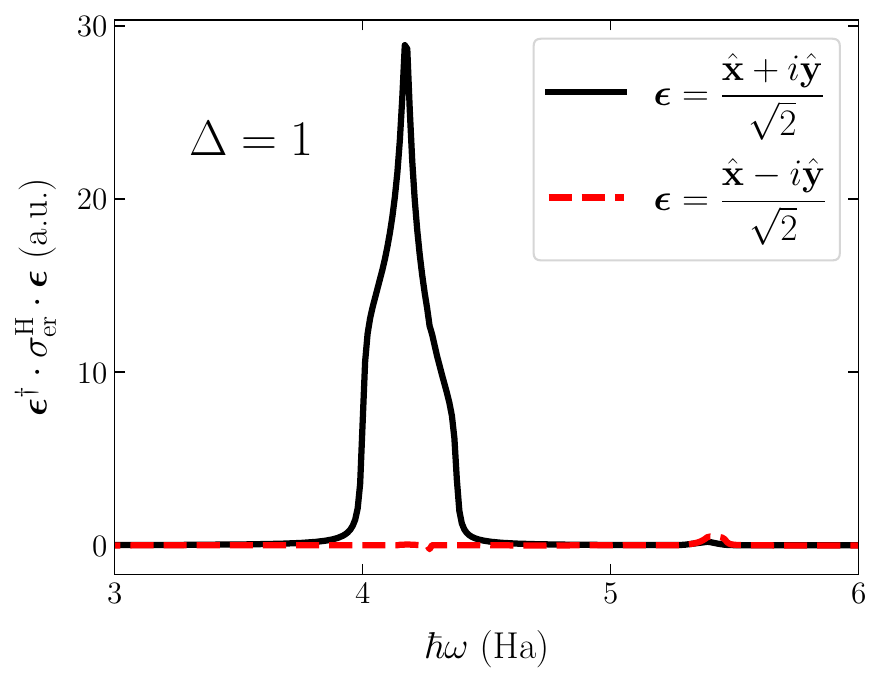}
\includegraphics[width=0.49\columnwidth]{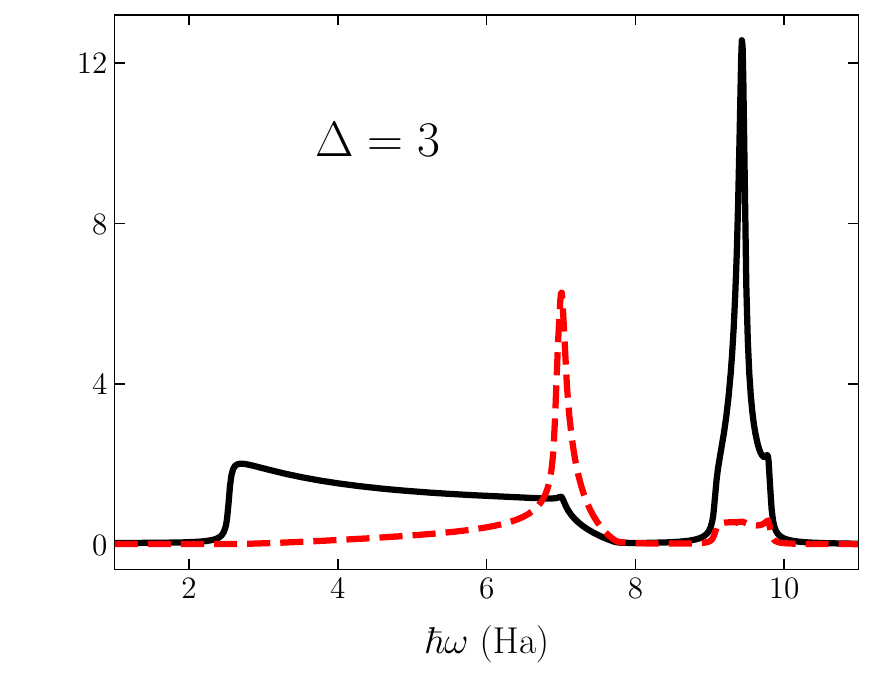}
\caption{Optical absorption spectra for circularly-polarized light in
  the tunable flat-band model with $\Delta = 1$ (left panel) and
  $\Delta =3$ (right panel). The plus sign in the polarization vector
  ${\boldsymbol\epsilon}$ corresponds to left circular polarization,
  or positive helicity, for light propagating along $+\hat{\bf z}$.}
\figlab{fig4}
\end{figure}

To address the weaker upper bound on
$n_{\rm e} \left| \langle \mm \rangle \right|$ (dotted line in
\fref{fig3}), we turn to the link to optical absorption.  According to
\sref{saturation}, that bound saturates when the system becomes
transparent at all frequencies for either left or right circular
polarization. This happens at $\Delta = 1$: the left panel of
\fref{fig4} shows strong absorption for left circular polarization
(but only from transitions to the middle band, consistent with the
earlier remarks on dipole transitions), and negligible optical
absorption for right circular polarization.  Instead, the right panel
shows that at $\Delta = 3$ there is significant absorption for both
polarizations, consistent with the unsaturated bound.

In summary, four conditions conspire to produce the near coincidence
at $\Delta = 1$ of the five curves in \fref{fig3}: nearly flat bands,
significant electric-dipole transitions from the filled band to the
first empty band only, a lack of sign changes in $\mm(\k)$ across the
BZ, and near-perfect optical transparency for one circular
polarization.  All four conditions are satisfied exactly in Landau
levels, where the bounds are fully saturated
(\sref{landau-levels-orb}).

\section{3D matrix-invariant inequalities}
\seclab{roy3d}

\subsection{General relations}

To extend the analysis of the previous section to 3D, take
$A = A' + iA''$ to be a $3 \times 3$ positive-semidefinite Hermitian
matrix. The principal minor test~\cite{prussing86} gives
\beq
\begin{aligned}
\det\, A \geq 0\,,\\
A_{xx}A_{yy} - |A_{xy}|^2 \geq 0\,, \;\;
A_{xx}A_{zz} - |A_{xz}|^2 \geq 0\,, \;\;
A_{yy}A_{zz} - |A_{yz}|^2 \geq 0\,,\\
A_{xx} \geq 0\,, \;\;
A_{yy} \geq 0\,, \;\;
A_{zz} \geq 0\,.
\eqlab{minors}
\end{aligned}
\eeq
To proceed, associate with the antisymmetric matrix $A''$ an axial
vector ${\bf A}''$ in the usual manner,
\beq
A''_\al = \frac{1}{2}\epsilon_{\alpha\beta\gamma} A''_{\beta\gamma}\,,
\eeq
and construct three positive-semidefinite $2\times 2$ matrices
$A'_\al$ by removing from $A'$ the $\al$-th row and the $\al$-th
column.  The principal invariants of these smaller symmetric matrices
are their traces,
\beq
\begin{aligned}
\tr\, A'_x &\equiv A'_{yy} + A'_{zz}
\left( = \lambda_2 + \lambda_3 \right)\,,
\\
\tr\, A'_y &\equiv A'_{zz} + A'_{xx}
\left( = \lambda_3 + \lambda_1 \right)\,,
\\
\tr\, A'_z &\equiv A'_{xx} + A'_{yy}
\left( = \lambda_1 + \lambda_2 \right)\,,
\end{aligned}
\eeq
and their determinants,
\beq
\begin{aligned}
\det\, A'_x &\equiv A'_{yy} A'_{zz} - A'_{yz} A'_{zy}
\left(
= \lambda_2 \lambda_3
\right)
\,,
\\
\det\, A'_y &\equiv A'_{zz} A'_{xx} - A'_{zx} A'_{xz}
\left(
= \lambda_1 \lambda_3
\right)
\,,
\\
\det\, A'_z &\equiv A'_{xx} A'_{yy} - A'_{xy} A'_{yx}
\left(
= \lambda_1 \lambda_2
\right)
\end{aligned}
\eeq
(the expressions in parentheses hold in the principal-axis frame of
$A'$). The relations in the second line of \eq{minors} can now be
brought to the same form as the 2D invariant relations in
\eq{ineqs-2D},
\beq
\left| A''_\al \right|
\leq
\sqrt{\det\, A'_\al}
\leq
\frac{1}{2} \tr\, A'_\al
\,,
\eqlab{ineqs-3D-comp}
\eeq
providing upper bounds on the magnitudes of the Cartesian components
of ${\bf A}''$.  To obtain bounds on its norm
$|{\bf A}''|=\sqrt{{\bf A}''\cdot{\bf A}''}$ --~the sole invariant of
$A''$~\cite{enwiki:1269822116}~-- align the $z$ axis with ${\bf A}''$
to find
\beq
\left| {\bf A}'' \right|
\leq
\sqrt{\det\, A'_z}
\leq
\frac{1}{2} \tr\, A'_z
\quad
\left( {\bf A}'' \parallel \hat{\bf z} \right)
\,.
\eqlab{ineqs-3D-a}
\eeq

One can establish additional relations involving scalar invariants of
the full $3\times 3$ matrix $A = A' + iA''$. Those invariants (six in
total) are the principal invariants of $A'$~\cite{holzapfel-book00},
\beq
\begin{aligned}
I_1 &= \tr\, A' =
\lambda_1 + \lambda_2 + \lambda_3 =
\frac{1}{2}
\left(
\tr\, A'_x +
\tr\, A'_y +
\tr\, A'_z  
\right)\,,
\\
I_2 &= \frac{1}{2}
\left[
\left(
\tr\, A' \right)^2 -
\tr \left( (A')^2
\right)
\right] =
\tr \left[(A')^{-1}\right] \det\, A'\\
&=
\lambda_1 \lambda_2 +
\lambda_1 \lambda_3 +
\lambda_2 \lambda_3 =
\det\, A'_x +
\det\, A'_y +
\det\, A'_z
\,,
\\
I_3 &= \det\, A' =
\lambda_1 \lambda_2 \lambda_3\,,
\eqlab{I1-I2-I3}
\end{aligned}
\eeq
together with the components $(A''_1,A''_2,A''_3)$ of ${\bf A}''$
along the principal axes $(\hat{\bf e}_1,\hat{\bf e}_2,\hat{\bf e}_3)$
of $A'$~\cite{enwiki:1269822116}.

Squaring the left inequality in \eq{ineqs-3D-comp} and summing over
$\al$ yields bounds on $\left| {\bf A}'' \right|$ involving $I_1$ and
$I_2$,
\beq
\left| {\bf A}'' \right|
\leq
\sqrt{
  \det\, A'_x +
  \det\, A'_y +
  \det\, A'_z
}
\leq
\frac{1}{\sqrt{3}} \tr\, A'\,;
\eqlab{ineqs-3D-b}
\eeq
the right inequality $I_2\leq I_1/\sqrt{3}$, proven in
Ref.~\cite{moriarty-elasticity71}, saturates when
$\lambda_1=\lambda_2=\lambda_3$.  These upper bounds on
$\left| {\bf A}'' \right|$ are generally less tight than the ones in
\eq{ineqs-3D-a}.

To obtain an inequality involving $I_3$, write the condition
$\det\, A \geq 0$ from \eq{minors} in the principal-axis frame,
\beq
\lambda_1 \left( A''_1 \right)^2 +
\lambda_2 \left( A''_2 \right)^2 +
\lambda_3 \left( A''_3 \right)^2
\leq
\lambda_1 \lambda_2 \lambda_3
\,.
\eqlab{ineqs-3D-c}
\eeq
If ${\bf A}''$ points along the principal axis $\hat {\bf e}_3$ chosen
as $\hat{\bf z}$, this becomes the left inequality in \eq{ineqs-3D-a}.

\subsection{Local and global inequalities}

When $A(\k) = T_p(\k)$, the left inequalities in
Eqs.~\eqref{eq:ineqs-3D-comp}, \eqref{eq:ineqs-3D-a} and
\eqref{eq:ineqs-3D-b} saturate in two-band models, and the one in
\eq{ineqs-3D-c} saturates in both two- and three-band models. These
properties are demonstrated in \aref{saturated}.

To obtain upper bounds on $\left| \boldsymbol{\mathcal A}'' \right|$,
align $\hat{\bf z}$ with $\boldsymbol{\mathcal A}''$ and then use the
triangle inequality
$\left| \boldsymbol{\mathcal A}'' \right| \leq \int_\k \left|
A''_z(\k) \right|$ to find
\beq
\left| \boldsymbol{\mathcal A}'' \right|
\leq
\int_\k \sqrt{\det\, A'_z(\k)}
\leq
\sqrt{\det\, {\mathcal A}'_z}
\leq
\frac{1}{2} \tr\, {\mathcal A}'_z
\quad
\left(
\boldsymbol{\mathcal A}'' \parallel \hat{\bf z}
\right)
\,,
\eqlab{global-3D-a}
\eeq
in direct analogy with the 2D inequalities in \eq{global-2D}. The
Minkowski inequality \eq{I2-minkowski} was used to go from the the
second to the third term, and \eq{ineqs-3D-comp} was used between the
third and the fourth.  As in the case of \eq{global-2D}, for
$A(\k) = T_p(\k)$ the first inequality saturates in two-band models if
$A''_z(\k)$ does not change sign across the BZ.

Another set of bounds follows from
$\left| \boldsymbol{\mathcal A}'' \right| \leq \int_\k \left| {\bf
  A}''(\k) \right|$,
\beq
\begin{aligned}
\left| \boldsymbol{\mathcal A}'' \right|
&\leq
\int_\k
\sqrt{
  \det\, A'_x(\k) +
  \det\, A'_y(\k) +
  \det\, A'_z(\k)
}\\
&\leq
\sqrt{
  \det\, {\mathcal A}'_x +
  \det\, {\mathcal A}'_y +
  \det\, {\mathcal A}'_z
}\\
&\leq
\frac{1}{\sqrt{3}}\, \tr\, \boldsymbol{\mathcal A}'\,.
\end{aligned}
\eqlab{global-3D-b}
\eeq
To go from the first line to the second, the following generalization
of the Minkowski inequality~\eqref{eq:minkowski} was
used~\cite{Marcus_Gordon_1971},
\beq
\sqrt{I_2(A)}+
\sqrt{I_2(B)}
\leq
\sqrt{I_2(A+B)}\,,
\eqlab{minkowski-gen}
\eeq
and from the second line to the third, \eq{ineqs-3D-b} was used.
Again, the bounds on $\left| \boldsymbol{\mathcal A}'' \right|$ in
\eq{global-3D-b} are generally less tight than their counterparts in
\eq{global-3D-a}.

\subsection{Metric-curvature inequalities}
\seclab{bounds-curv-3D}

When $A(\k)$ is the metric-curvature tensor, \eq{global-3D-a} gives
upper bounds on the norm of the Chern vector,
\beq
\frac{\left| {\bf K} \right|}{2\pi}
\leq
4\pi n_{\rm e}
\left<
\sqrt{\det g_z}
\right>
\leq
4\pi n_{\rm e}
\sqrt{\det \left< g_z \right>}
\leq
2\pi n_{\rm e}\, \tr \left< g_z \right>
\quad
\left(
{\bf K} \parallel \hat{\bf z}
\right)
\,,
\eqlab{bounds-K-a}
\eeq
where
\beq
g_z =
\begin{pmatrix}
g_{xx} & g_{xy}\\
g_{yx} & g_{yy}
\end{pmatrix}
\eeq
with $g_{xy} = g_{yx}$. These bounds are analogous to those in
\eq{ineqs-p0} on the magnitude of the Chern number in 2D.
\Eq{global-3D-b} gives the weaker bounds
\beq
\begin{aligned}
\frac{\left| {\bf K} \right|}{2\pi}
&\leq
4\pi n_{\rm e}
\left<
\sqrt{
  \det g_x +
  \det g_y +
  \det g_z
}
\right>\\
&\leq
4\pi n_{\rm e}
\sqrt{
  \det \left< g_x \right> +
  \det \left< g_y \right> +
  \det \left< g_z \right>
}\\
&\leq
\frac{4\pi}{\sqrt{3}} n_{\rm e}\, \tr \left< g \right>\,,
\eqlab{bounds-K-b}
\end{aligned}
\eeq
where
\beq
g_x =
\begin{pmatrix}
g_{yy} & g_{yz}\\
g_{zy} & g_{zz}
\end{pmatrix}\,,
\qquad
g_y =
\begin{pmatrix}
g_{zz} & g_{zx}\\
g_{xz} & g_{xx}
\end{pmatrix}
\eeq
with $g_{yz} = g_{zy}$ and $g_{zx} = g_{xz}$.

Similar relations can be readily established for the orbital
magnetization and for its optical part, starting from the
mass-magnetization and mass-moment tensors.

\subsubsection{Layered Haldane model}
\seclab{layered-haldane}

To illustrate the above relations, we adopt a 3D model from
Ref.~\cite{liu-nature22} consisting of AA-stacked Haldane-model layers
with one layer per cell. The in-plane and out-of-plane lattice
constants $a$ and $c$ are both set to $a_0$. The on-site energies on
the A and B sublattices are $E_0$ and $-E_0$, and the first- and
second-neighbor in-plane hoppings are $t_1=1$~Ha and $\pm it_2$, with
$t_2=1.2$~Ha.  Adjacent layers are coupled by vertical hoppings
$t_{\rm A} = 3$~Ha between $+E_0$ sites, and $t_{\rm B} = 0.5$~Ha
between $-E_0$ sites. The value of $E_0$ is treated as an adjustable
parameter.

In our calculations, the lower band is full and the upper band is
empty; since the top of the valence band goes above the bottom of the
conduction band for some values of $E_0$, this does not correspond to
filling the available states in order of increasing energy.

\begin{figure}
\centering
\includegraphics[width=0.65\columnwidth]{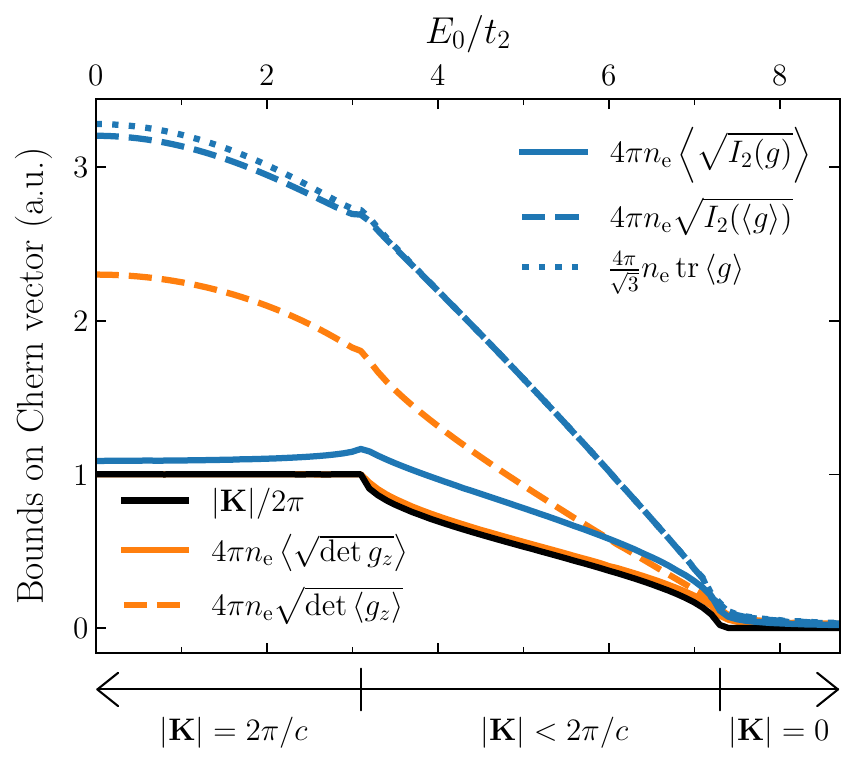}
\caption{Solid black line: Norm of the Chern vector ${\bf K}$ over
  $2\pi$ in a layered Haldane model, as a function of $E_0/t_2$ (the
  other model parameters are given in the main text). The lower band
  is treated as occupied.  The orange lines are the upper bounds in
  \eq{bounds-K-a} (two of them degenerate), and the blue lines are the
  upper bounds in \eq{bounds-K-b}. The gapped phases with
  $|{\bf K}|=2\pi/c$ and $|{\bf K}|=0$, and the intermediate gapless
  phase with nonquantized $|{\bf K}|$, are indicated on the bottom.}
\figlab{fig5}
\end{figure}

At $E_0=0$, the system is in a gapped phase with a quantized Chern
vector ${\bf K} = (0,0,2\pi/c)$.  At $E_0/t_2\simeq 3.1$, it enters a
gapless phase where initially the bands touch at a single point
located at the intersection between the $k_z = 0$ plane and a vertical
BZ edge.  As $E_0/t_2$ increases further, the band touching splits
into a pair of isolated Weyl nodes located on either side of the
$k_z=0$ plane; the Chern vector becomes ${\bf K} = (0,0,K_z)$, with
$K_z$ decreasing from $2\pi/c$ as the nodes drift apart. At
$E_0/t_2\simeq 7.3$ the two nodes meet again and annihilate on the
$k_z= \pm \pi/c$ plane, and the system enters a gapped phase with
${\bf K}={\bf 0}$.

In \fref{fig5}, the solid black line shows $|{\bf K}|/2\pi$
versus $E_0/t_2$, and the other lines show various upper bounds on
that quantity. The first bound in \eq{bounds-K-a} (solid orange line)
remains very tight over the entire range of the figure.  The dashed
orange line gives the other two bounds in that equation; those bounds
are degenerate thanks to the rotational symmetry of the model, which
makes the tensor $\langle g_z \rangle$ isotropic.  The three blue
lines represent the bounds in \eq{bounds-K-b}, which for the most part
remain significantly less tight than their counterparts in
\eq{bounds-K-a}.

\section{Gap inequalities and Cauchy-Schwarz  inequalities}
\seclab{spectral}

So far, we have considered relations between scalar invariants of
positive-semidefinite Hermitian matrices. These include inequalities
between sum-rule quantities on the same row but in different columns
of Table~\ref{tab:sum-rules}, such as the metric-curvature and
mass-moment inequalities.

We now turn to relations between sum-rule quantities located on
different rows of the left column of Table~\ref{tab:sum-rules}.  In
the following, two types of relations are discussed: gap inequalities,
and Cauchy-Schwarz inequalities.  By combining them with
matrix-invariant inequalities, one obtains relations between
quantities on different rows and different columns of
Table~\ref{tab:sum-rules}.

\subsection{Gap inequalities}
\seclab{gap-ineqs}

Consider the sum-rule tensor $S_p$ defined by \eq{sum-rule}.  Since
it is positive semidefinite, we have
\beq
S_p^{\boldsymbol\epsilon} \equiv
{\boldsymbol\epsilon}^\dagger
\cdot
S_p
\cdot
{\boldsymbol\epsilon} \geq 0
\quad
\text{for all } {\boldsymbol \epsilon} \in {\mathbbm C}^d\,;
\eqlab{Sp-posdef}
\eeq
since interband absorption of light with arbitrary polarization
${\boldsymbol \epsilon}$ is non-negative at all frequencies and
vanishes below the optical gap $\Eg$ (defined as the threshold energy
for interband absorption), we also have, by \eq{Sp-posdef-int},
\beq
S_{p-q}^{\boldsymbol\epsilon}
\leq
\left( \frac{\hbar}{E_{\rm g}} \right)^q
S_p^{\boldsymbol\epsilon}
\quad
\text{for $q>0$}\,.
\eeq
Equivalently,
\beq
{\boldsymbol\epsilon}^\dagger
\cdot
\left[
S_p -
\left( \frac{E_{\rm g}}{\hbar} \right)^q
S_{p-q}
\right]
\cdot
{\boldsymbol\epsilon}
\geq
0
\quad
\text{for $q>0$}
\,,
\eqlab{gap-posdef}
\eeq
which means that the tensor inside square brackets is also positive
semidefinite. These inequalities saturate when the absorption spectrum
is concentrated at $\Eg$, and they constitute stricter versions of the
inequality in \eq{Sp-posdef}.\footnote{Based on this observation, one
  can deduce tighter versions of the matrix-invariant inequalities of
  \srefs{roy}{roy3d} applied to $S_p \propto \int_\k T_p(\k)$. To that
  end, simply apply those same inequalities to the ``less positive''
  tensor inside square brackets in \eq{gap-posdef}.} For linear
polarization along direction $\al$, they reduce
to~\cite{souza-prb00,onishi-prx24,souza-scipost25}
\beq
S'_{p-q,\al\al}
\leq
\left( \frac{\hbar}{E_{\rm g}} \right)^q
S'_{p,\al\al}
\quad
\text{for $q>0$}
\,.
\eqlab{gap-ineqs-linear}
\eeq
Since $\Eg$ is at least as large as the minimum direct gap, the
inequalities continue to hold when $\Eg$ is replaced by that
gap~\cite{onishi-prx24} (in all our examples, the optical gap
coincides with the minimum direct gap).  By adjusting $p$ and $q$, gap
relations between different quantities on the left column of
Table~\ref{tab:sum-rules} can be
deduced~\cite{souza-prb00,aebischer-prl01,martin-book04,onishi-prx24,komissarov-natcomms24,onishi-prb24,verma-pnas25,souza-scipost25,onishi-prr25}.
Similar relations are discussed in Ref.~\cite{traini-epj96} for atoms.

\subsubsection{Localization length bounded by the inverse gap}

As an example, consider \eq{gap-ineqs-linear} with $p=q=1$,
\beq
S'_{0,\al\al}
\leq
\frac{\hbar}{E_{\rm g}} S'_{1,\al\al}\,.
\eqlab{gap-pq-1}
\eeq
Consulting Table~\ref{tab:sum-rules}, one finds
\beq
\langle g \rangle_{\al\al}
\leq
\frac{\hbar^2}{2\Eg} \left< m^{-1}_{\rm er} \right>_{\al\al}
\leq
\frac{\hbar^2}{2\Eg} \left< m^{-1}_* \right>_{\al\al}\,.
\eqlab{metric-gap}
\eeq
In insulators $\langle g \rangle_{\al\al}$ equals the squared
localization length, and the second inequality saturates. For an
insulator with a microscopic Hamiltonian, \eq{metric-gap}
reads~\cite{souza-prb00}
\beq
\ell^2_{\al\al}
\leq
\frac{\hbar^2}{2m_{\rm e}\Eg}
\Leftrightarrow
\left(
\frac{\ell_{\al\al}}{a_0}
\right)^2
\leq
\frac{{\rm Ha}}{2\Eg}\,.
\eqlab{loc-gap-micro}
\eeq

\subsection{Cauchy-Schwarz inequalities}
\seclab{cauchy-schwarz}

Consider now the Cauchy-Schwarz inequality
\beq
\left|
\int_0^\infty d\w\, f_1(\w) f_2(\w)
\right|^2
\leq
\int_0^\infty d\w\, \left| f_1(\w) \right|^2
\int_0^\infty d\w\, \left| f_2(\w) \right|^2
\,,
\eqlab{cauchy-schwarz}
\eeq
and choose
\beq
\begin{aligned}
f_1(\w) &= \w^{p/2}
\sqrt{
  {\boldsymbol\epsilon}^\dagger \cdot
  {\boldsymbol\sigma}^{\rm H}(\w) \cdot
  {\boldsymbol\epsilon}
}\,,\\
f_2(\w) &= \w^{(p-2)/2}
\sqrt{
  {\boldsymbol\epsilon}^\dagger \cdot
  {\boldsymbol\sigma}^{\rm H}(\w) \cdot
  {\boldsymbol\epsilon}
}
\end{aligned}
\,.
\eqlab{f-g}
\eeq
Since ${\boldsymbol\sigma}^{\rm H}(\w)$ is positive semidefinite, the
quantity inside the square root is non-negative. Using the sum
rule~\eqref{eq:sum-rule}, one obtains
\beq
S^{\boldsymbol\epsilon}_p
\leq
\sqrt{S^{\boldsymbol\epsilon}_{p+1} S^{\boldsymbol\epsilon}_{p-1}}\,.
\eqlab{S-cauchy}
\eeq
Such relations have also been discussed for atoms~\cite{traini-epj96}.
Like the gap inequalities, they saturate in the limit of an infinitely
narrow absorption spectrum.

\subsubsection{Localization length bounded by the susceptibility}

\begin{figure}
\centering
\includegraphics[width=0.65\columnwidth]{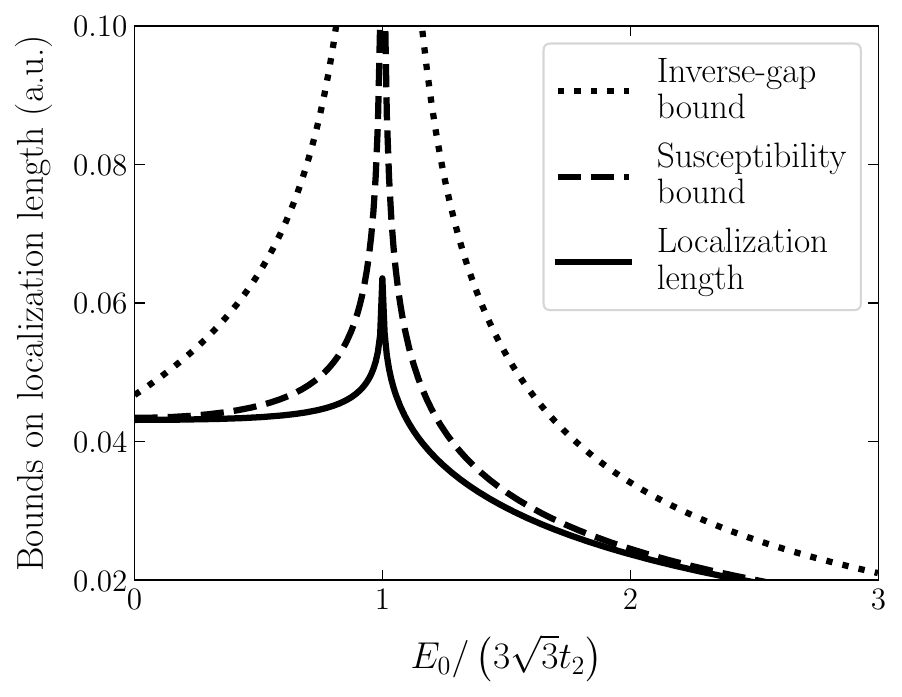}
\caption{Upper bounds on the electronic localization length in the 2D
  Haldane model, with the lower band occupied and the same choice of
  parameters as in \sref{haldane-2D}.  Solid line: localization length
  $\ell$. Dotted line: weak upper bound set by the inverse gap via
  \eq{metric-gap}.  Dashed line: strong upper bound set by the
  clamped-ion electric susceptibility via \eq{loc-chi}. All three
  quantities diverge at the boundary between the topological and
  trivial insulating phases, where the direct gap closes.}
\figlab{fig6}
\end{figure}

For our purposes, \eq{S-cauchy} is most useful with $p=0$ and with
${\boldsymbol\epsilon}$ a unit vector along direction~$\alpha$,
\beq
S'_{0,\al\al}
\leq
\sqrt{S'_{1,\al\al} S'_{-1,\al\al}}\,.
\eqlab{S-cauchy-lin}
\eeq
Referring to Table~\ref{tab:sum-rules} and specializing to insulators,
this becomes an upper bound on the localization length set by the
clamped-ion electric
susceptibility~\cite{verma-pnas25,souza-scipost25,onishi-prr25},
\beq
\ell^2_{\al\al}
\leq
\frac{\hbar}{2|e|}
\sqrt{
  \frac{ \langle m^{-1}_* \rangle_{\al\al} \epsilon_0 \chi_{\al\al}(0) }
  {n_e}
}\,.
\eqlab{loc-chi}
\eeq
This bound is tighter than the one of
\eq{loc-gap-micro}~\cite{souza-scipost25,onishi-prr25}, as illustrated
in \fref{fig6} for the 2D Haldane model.

\subsection{Bounds on the energy gap}

The combination of the inverse-gap bound~\eqref{eq:loc-gap-micro} on
$\ell^2$ with the metric-curvature inequalities~\eqref{eq:ineqs-p0}
and~\eqref{eq:bounds-K-a} yields upper bounds on the optical gap set
by the Chern invariants~\cite{onishi-prx24},
\beq
\begin{aligned}
\Eg
&\leq
2\pi\hbar^2
\frac{n_{\rm e}/m_{\rm e}}{|C|}\
\,
\Leftrightarrow
\frac{\Eg}{\rm Ha}
\leq
\left( \frac{a_0}{r_{\rm s}} \right)^2 \frac{2}{|C|}
\quad\quad\,
\text{(2D)}\,,
\\
\Eg
&\leq
2\pi\hbar^2
\frac{n_{\rm e}/m_{\rm e}}{\left| {\bf K} \right|/2\pi}
\Leftrightarrow
\frac{\Eg}{\rm Ha}
\leq
\left( \frac{a_0}{r_{\rm s}} \right)^3
\frac{3\pi}{a_0 \left| {\bf K} \right|}
\quad
\text{(3D)}\,.
\end{aligned}
\eqlab{C-K-gap}
\eeq
The radius $r_{\rm s}$ is defined by $1/n_{\rm e}=\pi r_{\rm s}^2$ in
2D, and by $1/n_{\rm e}=(4\pi/3) r_{\rm s}^3$ in 3D.

Before applying it to the layered Haldane model, let us examine more
closely the 3D version of the Chern bound. The form applicable to
tight-binding models is
\beq
\Eg
\leq
2\pi\hbar^2
\frac{n_{\rm e}/\left< m_{\rm er} \right>_\perp}{\left| {\bf K} \right|/2\pi}
\leq
2\pi\hbar^2
\frac{n_{\rm e}/\left< m_* \right>_\perp}{\left| {\bf K} \right|/2\pi}
\,,
\eqlab{K-gap-eff}
\eeq
assuming isotropic inverse-mass tensors on the plane perpendicular to
${\bf K}$.  The notation is
\beq
\langle m_*\rangle_\perp =
1/\left< m^{-1} \right>_{xx} = 1/\left< m^{-1} \right>_{yy}\,,
\eeq
with $x$ and $y$ the in-plane directions.  In any gapped phase,
$\left< m_{\rm er} \right>_\perp = \left< m_* \right>_\perp$. In the
layered Haldane model, this equality holds in the gapless phase as
well: since we treat the lower band as occupied, that phase mimics an
ideal Weyl semimetal with a point-like Fermi surface, for which the
Drude weight
$D_\perp \propto n_{\rm e}/\left< m_* \right>_\perp - n_{\rm e}/\left<
m_{\rm er} \right>_\perp$ vanishes.

In Ref.~\cite{souza-scipost25}, a sequence of three upper bounds on
the optical gap of a generic insulator was obtained by combining the
Cauchy-Schwarz inequality~\eqref{eq:S-cauchy-lin} with several
instances of the gap inequality~\eqref{eq:gap-ineqs-linear}.  Those
bounds can be chained with the Chern bound~\eqref{eq:K-gap-eff} as
follows,
\beq
\begin{aligned}
\Eg
&\leq
\frac{2e^2 n_{\rm e} \ell^2_\perp}{\epsilon_0 \chi_\perp(0)}
\leq
\hbar |e|
\sqrt{
  \frac{n_{\rm e}}{\left< m_* \right>_\perp \epsilon_0 \chi_\perp(0)}
}\\
&\leq
\frac{\hbar^2}{2 \left< m_* \right>_\perp \ell^2_\perp}
\leq
2\pi\hbar^2
\frac{n_{\rm e}/\left< m_* \right>_\perp}{\left| {\bf K} \right|/2\pi}\,.
\end{aligned}
\eqlab{gap-bounds}
\eeq
The outermost inequality is \eq{K-gap-eff} written for insulators, and
the second line corresponds to the metric-curvature inequality of
\eq{bounds-K-a}. The fact that the Chern bound is the weakest in a
sequence of upper bounds on $\Eg$ suggests that it is not tight in
general.  Indeed, the three intermediate bounds only approach $\Eg$
when the entire absorption spectrum is concentrated at
$\Eg$~\cite{souza-scipost25}.  Moreover, the last inequality only
saturates under very specific conditions, as discussed for the 2D case
in \sref{saturation}.

\Eq{gap-bounds} may be written concisely as
\beq
\Eg \leq E^2_{\rm P}/E_\ell \leq E_{\rm P} \leq E_\ell \leq E_{\rm C}\,,
\eqlab{gap-bounds-energies}
\eeq
where the subscript $\perp$ has been omitted.  $E_{\rm P}$ and
$E_\ell$ are the Penn and localization gaps representing different
frequency-weighted averages of the absorption spectrum of
linearly-polarized light~\cite{martin-book04,souza-scipost25}, and
$E_{\rm C}$ is the Chern bound.

\subsubsection{Layered Haldane model}

\begin{figure}
\centering
\includegraphics[width=0.65\columnwidth]{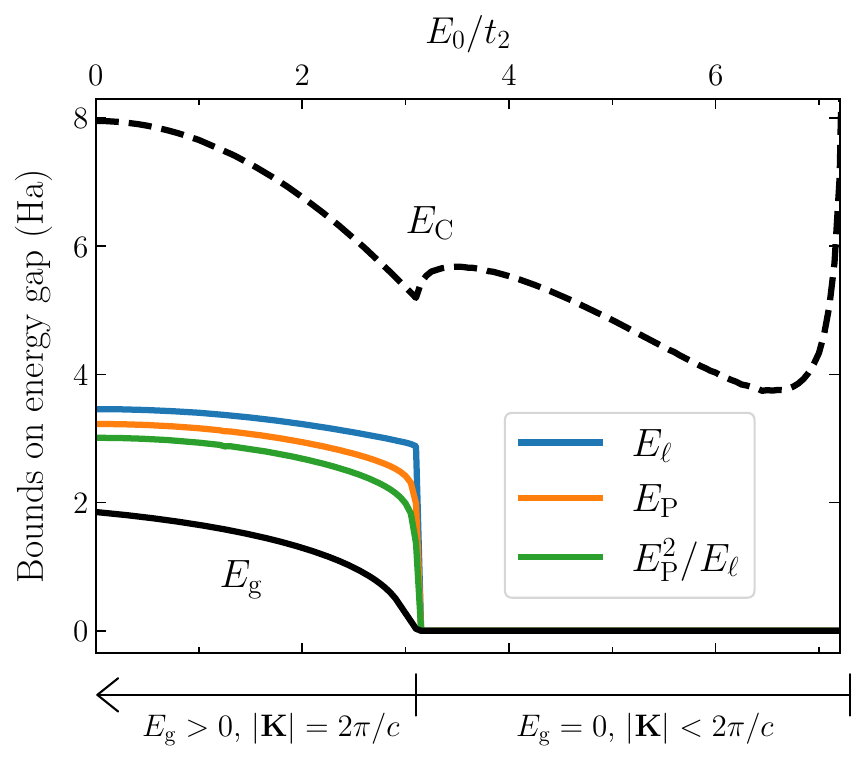}
\caption{Solid black line: minimum direct gap $\Eg$ of the layered
  Haldane model of \sref{layered-haldane} in the gapped and gapless
  phases with nonzero Chern vector ${\bf K}$.  Dashed black line:
  outermost bound on $\Eg$ in \eq{gap-bounds-energies}, set by the
  inverse magnitude of ${\bf K}$. The Chern vector goes to zero on the
  right end of the plot (see \fref{fig5}), producing the divergence of
  the bound. The solid colored lines represent the intermediate
  quantities in \eq{gap-bounds-energies}, which vanish in the gapless
  phase.}
\figlab{fig7}
\end{figure}

All five terms in \eq{gap-bounds-energies} are plotted in \fref{fig7}
for the layered Haldane model in the gapped and gapless phases with
nonzero Chern vector.  In the gapped phase, all four bounds on $\Eg$
are far from saturation.  In the gapless phase where $\Eg=0$, the
intermediate bounds go to zero because of the divergence of
$\ell^2_\perp$ and the even faster divergence of $\chi_\perp(0)$,
while the Chern bound remains large, diverging at the boundary with
the trivial insulating phase.  In a typical ferromagnetic metal $\Eg$
would remain nonzero, and the two forms of the Chern bound given in
\eq{K-gap-eff} would become nondegenerate.

\subsubsection{2D flat-band model}

\begin{figure}
\centering
\includegraphics[width=0.49\columnwidth]{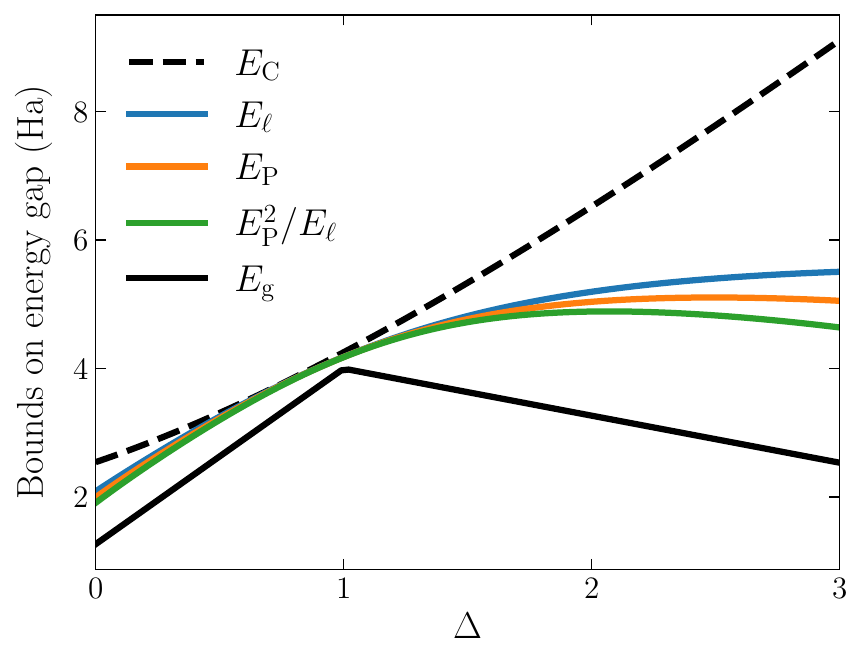}
\includegraphics[width=0.49\columnwidth]{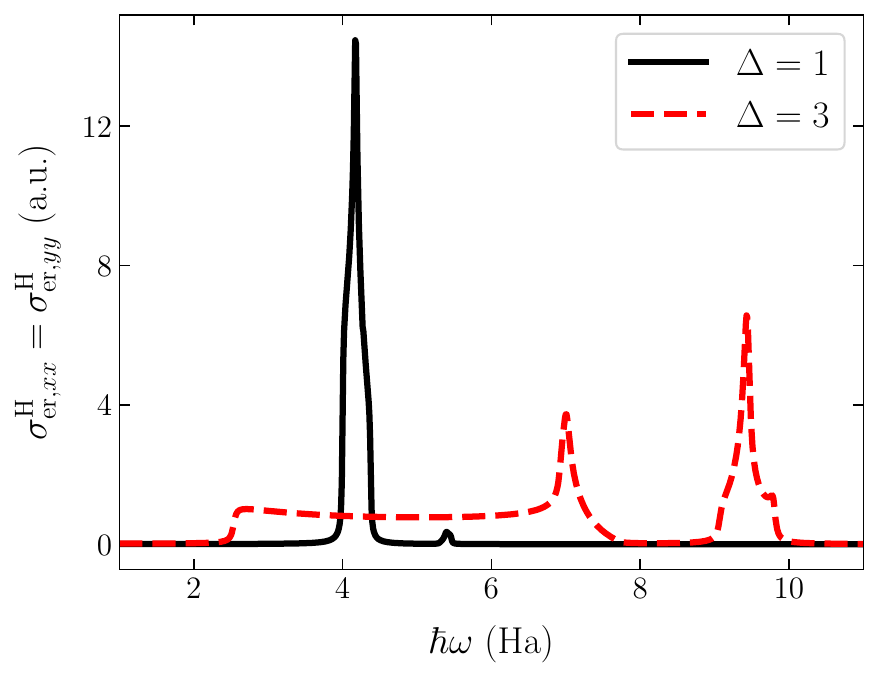}
\caption{Left: same quantities as in \fref{fig7}, but for
  the 2D flat-band model of \sref{flatband-magnetization}, plotted
  versus the flatness parameter $\Delta$. Right: absorption spectrum
  for linearly-polarized light at $\Delta = 1$ (narrow bands) and
  $\Delta = 3$ (wide bands).}
\figlab{fig8}
\end{figure}

The conditions under which the four upper bounds on the energy gap of
a Chern insulator saturate are nearly met in the flat-band model of
\sref{flatband-magnetization} at $\Delta = 1$, as shown in the left
panel of \fref{fig8}. To shed light on the near saturation of the
three non-Chern bounds, the right panel displays the absorption
spectra of linearly-polarized light at both $\Delta=1 $ and
$\Delta = 3$.

At $\Delta = 3$, the spectrum consists of a broad plateau with a sharp
peak at the upper end from transitions between the lower and middle
bands, and a narrow peak from transitions between the lower and upper
bands; these features are consistent with the band dispersions in the
right panel of \fref{fig2}.  Since the spectrum is quite broad, the
three non-Chern bounds on $\Eg$ are not well saturated.  The
additional separation $E_{\rm C} - E_\ell$ that renders the Chern
bound even less saturated is caused by the sizable absorption of both
left- and right-circularly-polarized light (right panel in
\fref{fig4}).

At $\Delta = 1$, the absorption of linearly-polarized light is
strongly peaked at the energy separation between the weakly-dispersive
lower and middle bands, with only a tiny peak at higher frequencies
from transitions to the upper band, in agreement with the approximate
selection rules discussed in \sref{flatband-magnetization}; this
narrow spectrum results in the near saturation of the non-Chern
bounds.  The separation $E_{\rm C} - E_\ell$ is also quite small
thanks to the almost perfect transparency for right circular
polarization (left panel in \fref{fig4}), producing the near
saturation of the Chern bound.

\subsection{Topological bound on the susceptibility}
\seclab{bound-chi}

In combination with the metric-curvature
inequalities~\eqref{eq:ineqs-p0} and~\eqref{eq:bounds-K-a},
\eq{loc-chi} yields lower bounds on the electric susceptibilities of
2D and 3D Chern insulators; for microscopic Hamiltonians and in-plane
isotropic $\chi(0)$ and $\ell^2$ tensors, those bounds take the simple
forms
\beq
\begin{aligned}
\chi(0)
&\geq
\frac{C^2}{\pi a_0 n_{\rm e}}
\quad \quad\;\;\;
\text{(2D)}\,,
\\
\chi_\perp(0)
&\geq
\frac{(\left| {\bf K} \right|/2\pi)^2}{\pi a_0 n_{\rm e}}
\quad
\text{(3D)}\,.
\end{aligned}
\eqlab{chi-C-K}
\eeq
The susceptibility is dimensionless in 3D, while in 2D it has units of
length.  Viewed as upper bounds on $|C|$ and $|{\bf K}|$, these
relations are tighter than the ones in \eq{C-K-gap}: the 3D relation
in \eq{chi-C-K} corresponds to $E_{\rm P} \leq E_{\rm C}$ in
\eq{gap-bounds-energies}, while the one in \eq{C-K-gap} is
$\Eg \leq E_{\rm C}$.

\subsubsection{Landau levels}

As a first example, consider the 2D free electron gas in a magnetic
field already discussed in \sref{landau-levels-orb}. In the integer
quantum Hall state with the $\nu$ lowest Landau levels filled, the 2D
inequality in \eq{chi-C-K} saturates (\aref{LL}). Since the magnitude
of the Chern number equals $\nu$ (\aref{LL}), the electron density,
$n_{\rm e} = (\nu/h)|eB|$, can be written as
$n_{\rm e} = m_{\rm e}\w_{\rm c} |C|/h$ with
$\w_{\rm c}=|eB|/m_{\rm e}$ the cyclotron frequency. Hence,
\beq
\chi(0) =
\frac{C^2}{\pi a_0 n_{\rm e}} =
\frac{C^2}{\pi}
\left(
\frac{e^2 m_{\rm e}}{4\pi \epsilon_0 \hbar^2}
\right)
\frac{h}{m_{\rm e} \w_{\rm c}|C|} =
\frac{1}{2\pi \epsilon_0}
\frac{e^2 |C|}{ \hbar \w_{\rm c}}\,,
\eeq
which agrees with the expression obtained in
Ref.~\cite{komissarov-natcomms24} by different means. The present
derivation shows how it emerges from the saturation of a general bound
relation. 

The addition of a weak substrate potential changes neither the
electron density nor the Chern number. The only change to the above
relation for a non-flat Landau-level system at integer filling is that
it becomes an inequality,
\beq
\chi(0)
\geq
\frac{1}{2\pi \epsilon_0}
\frac{e^2 |C|}{ \hbar \w_{\rm c}}\,,
\eeq
where isotropy of $\chi(0)$ in the presence of the substrate is
assumed for simplicity.  Again, this result is a special case of the
2D bound relation in \eq{chi-C-K}.  As seen in that equation, for
fixed density the dependence on the Chern invariant is quadratic, not
linear.

\subsubsection{2D Haldane and flat-band models}

\begin{figure}
\centering
\includegraphics[width=0.49\columnwidth]{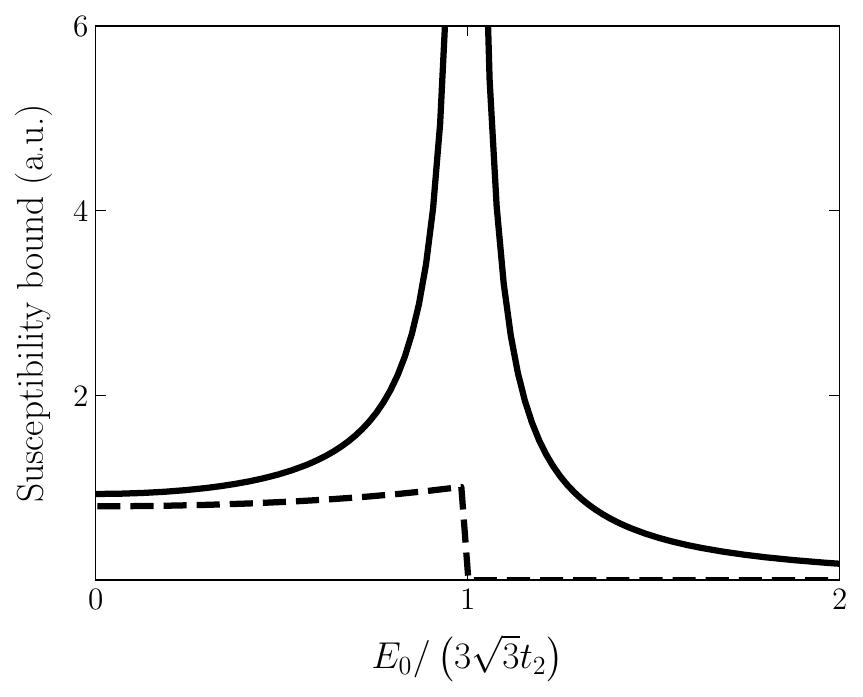}
\includegraphics[width=0.472\columnwidth]{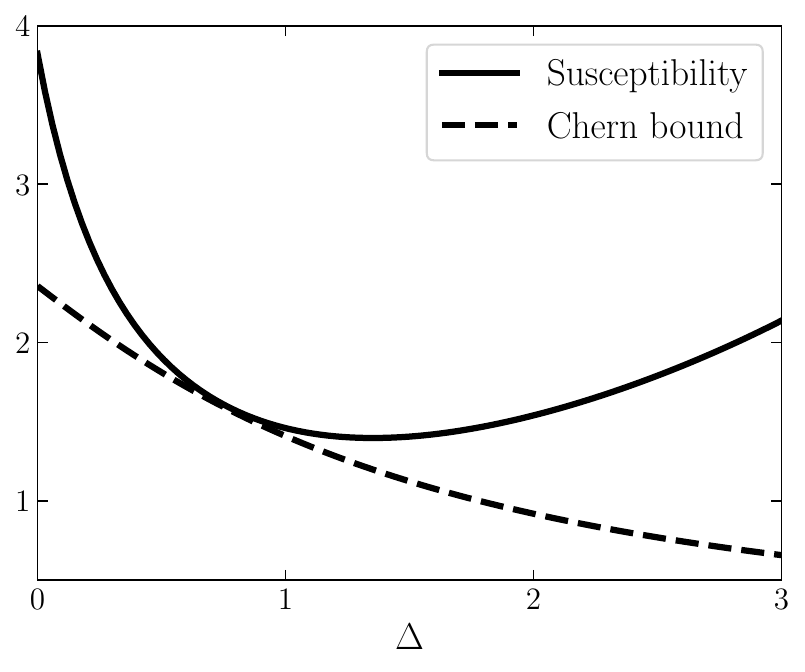}
\caption{Clamped-ion electric susceptibility (solid line),
    and Chern bound on the susceptibility (dashed line).  Left: 2D
    Haldane model of \sref{haldane-2D}. Right: 2D flat-band model of
    \sref{flatband-magnetization}.}
  \figlab{fig9}
\end{figure}

For a tight-binding model with in-plane isotropy, the Chern bound on
the susceptibility is still given by \eq{chi-C-K}, but with $a_0$
replaced by a renormalized Bohr radius defined as
\beq
a_\perp^* =
\frac{4\pi \epsilon_0 \hbar^2}{e^2 \left< m_* \right>_\perp}\,.
\eqlab{a0-eff}
\eeq
The resulting bound is displayed in the left panel of \fref{fig9} for
the 2D Haldane model.  At the boundary between the two gapped phases
the susceptibility diverges, while the lower bound drops
discontinuously from a finite value on the $|C|=1$ phase to zero on
the $C=0$ phase. Far from the phase boundary, the bound on the
susceptibility becomes progressively more saturared.

Like the bounds on the orbital magnetization (\fref{fig3}) and on the
energy gap (\fref{fig8}), the Chern bound on the susceptibility gets
nearly saturated for the flat-band model, as shown in the right panel
of \fref{fig9}. The near saturation occurs slightly below
$\Delta = 1$, as in the case of the global metric-curvature
bounds~\cite{ozawa-prb21}.

\section{Conclusions}
\seclab{conclusions}

This work explored bound relations on quantum-geometric electronic
properties of magnetic crystals, and their interplay with measurable
responses functions: optical absorption, anomalous Hall conductivity,
and clamped-ion electric susceptibility.  Previous works were extended
by a careful consideration of the Chern invariant in 3D and by
treating partially filled bands in metals, and certain results in the
literature were shown to be special cases of more general relations.
Although the discussion was framed in the mean-field language of band
theory, the various bound relations, being rooted on fundamental
ground-state quantities and optical sum rules, remain valid for
correlated and disordered systems.

Several of the bound relations were illustrated for 2D and 3D model
systems: Landau levels, Chern insulators, and Weyl semimetals.  The
conditions under which those relations become tight in a 2D model with
tunable flat bands were analyzed in terms of the optical absorption
spectrum for linear and circular polarization. In addition to band
flatness, several other factors were found to play a role in achieving
near saturation of the bound relations: optical selection rules,
strong magnetic circular dichroism, and the absence of sign changes in
the Berry curvature and orbital moment across the BZ.

The study of sum rules and bound relations in real materials via
first-principles calculations~\cite{ghosh-sciadv24} and experimental
measurements~\cite{balut-prb25} is in the early stages, and
constitutes a promising research direction.  Some of the relations
established herein provide new opportunities, particularly in
connection with flat-band Chern materials and generalized Landau
systems~\cite{ledwith-prb23,fujimoto-prl25,liu-prx25}.  For example,
the tight Chern bound on the dc electric susceptibility, involving
directly measurable quantities --~electron density, Chern number, and
susceptibility~-- could help probe topological flat-band systems where
the metric-curvature inequality saturates.

\section*{Acknowledgments}
Stimulating discussions with Jos\'e Lu\'is Martins and Cheol-Hwan Park
are gratefully acknowledged.

\paragraph{Funding information}
This work was supported by Grant No.  PID2021-129035NB-I00 funded by
MCIN/AEI/10.13039/501100011033 and by ERDF/EU, and by the Marie
Skłodowska-Curie grant agreement No. 101206626.

\begin{appendix}
\numberwithin{equation}{section}

\section{Derivation of \eq{T-tr} for $T_{p \geq 0}(\k)$}
\seclab{T-tr-der}

Take \eq{T-geom} for $T_p(\k)$, and expand it for $p \geq 0$ using the
binomial theorem to find
\beq
T^\ab_{p \geq 0} =
\sum_{l=0}^p\, (-1)^l
\begin{pmatrix}
p\\
l
\end{pmatrix}
\sum_{n\in\Fk}\,
\me{\partial_\al u\bnk}{Q_\k H_\k^{p-l} Q_\k}{\partial_\be u\bnk}
\enk^l\,,
\eqlab{T-bin}
\eeq
where $Q_\k \equiv \one - P_\k$.  To show that this is equivalent to
\eq{T-tr}, note the identity
\beq
\partial_\al P_\k =
P_\k\left(\partial_\al P_\k\right)Q_\k +
Q_\k\left(\partial_\al P_\k\right)P_\k\,,
\eqlab{delP}
\eeq
which states that the only nonvanishing matrix elements of
$\partial_\al P$ are those that connect $\Fk$ to its complement
space. \Eq{delP} is easily proven by writing the projectors explicitly
in terms of kets and bras. This identity can be used to manipulate the
following electronic trace,

\beq
\begin{aligned}
\Tr
\left[
P_\k
\left( \partial_\alpha P_\k \right)
H_\k^i
\left( \partial_\beta P_\k \right)
H_\k^j
\right] &= \Tr
\left[
P_\k
\left( \partial_\alpha P_\k \right)
Q_\k\,
H_\k^i
\left( \partial_\beta P_\k \right)
H_\k^j
\right]\\
&=\Tr
\left[
P_\k
\left( \partial_\alpha P_\k \right)
\left( Q_\k\,H_\k^i\,Q_\k \right)
\left( \partial_\beta P_\k \right)
H_\k^j
\right]
\,,
\end{aligned}
\eeq
where the last step used $Q^2=Q$ and $[H,Q]=0$. Writing the trace
explicitly with $i=p-l$ and $j=l$ gives
\beq
\Tr
\left[
P_\k
\left( \partial_\alpha P_\k \right)
H_\k^{p-l}
\left( \partial_\beta P_\k \right)
H_\k^l
\right] = \sum_{n\in\Fk}\,
\me{\partial_\al u\bnk}{Q_\k H_\k^{p-l} Q_\k}{\partial_\be u\bnk}
\enk^l\,,
\eqlab{T-gen}
\eeq
which combined with \eq{T-bin} yields \eq{T-tr}.

\section{$f$- sum rule and effective masses}
\seclab{f-sum-rule}

Using \eq{T1} for $T_p(\k)$, the real part of the interband sum rule
in \eq{sum-rule} with $p=1$ becomes
\beq
\frac{2}{\pi}\int_0^\infty d\w\,
\Re\,\sigma^{\rm S}_{{\rm er},\ab}(\w) =
e^2 n_{\rm e} \left< m^{-1}_{\rm er} \right>_\ab\,,
\eqlab{f-er}
\eeq
where $m^{-1}_{\rm er}$ is the interband inverse mass of \eq{mass-er}.
Intraband absorption contributes the additional amount
\beq
\frac{2}{\pi}\int_0^\infty d\w\,
\Re\,\sigma^{\rm S}_{{\rm ra},\ab}(\w) =
\frac{e^2}{\hbar^2}\sum_n \int_\k
\left( \partial_\al \enk \right)
\left( \partial_\be \enk \right)
\left(
-\frac{\partial f}{\partial \epsilon}
\right)_{\epsilon = \enk}\,,
\eqlab{f-ra-surf}
\eeq
with $f(\epsilon)$ the Fermi-Dirac distribution function.  Converting
from a Fermi-surface to a Fermi-sea integral via an integration by
parts, one obtains at zero temperature
\beq
\frac{2}{\pi}\int_0^\infty d\w\,
\Re\,\sigma^{\rm S}_{{\rm ra},\ab}(\w) =
e^2 n_{\rm e} \left< m^{-1}_{\rm ra} \right>_\ab\,,
\eqlab{f-ra-sea}
\eeq
with $m^{-1}_{\rm ra}$ the intraband inverse mass of
\eq{mass-ra}. Since $k$-space periodicity was invoked to discard the
boundary terms, \eq{f-ra-sea} is not valid for low-energy continuum
models, as illustrated below for graphene.

The tensors $\left< m^{-1}_{\rm er} \right>$ and
$\left< m^{-1}_{\rm ra} \right>$ --~and hence the Drude
weight~\eqref{eq:drude}~-- inherit the positive semidefiniteness of
$\sigma^{\rm H}_{\rm er}(\w)$ and $\sigma^{\rm H}_{\rm ra}(\w)$.  But
while $m^{-1}_{\rm er}(\k)$ is locally positive semidefinite
(\sref{posdef}), $m^{-1}_{\rm ra}(\k)$ only becomes positive
semidefinite after averaging over the Fermi sea; instead, the
Fermi-surface expression in \eq{f-ra-surf} is locally positive
semidefinite.

Combining \eqs{f-er}{f-ra-sea} and introducing the optical inverse
mass via \eq{masses} leads to the oscillator-strength sum rule (or
$f$-sum rule)
\begin{align}
\frac{2}{\pi}\int_0^\infty d\w\,\Re\,\sigma^{\rm S}_\ab(\w)
&=e^2 n_{\rm e} \left< m^{-1}_{*} \right>_\ab\,,
\eqlab{f-sum}
\end{align}
where $\sigma = \sigma_{\rm er} + \sigma_{\rm ra}$.  This form of the
$f$-sum rule remains valid for nonlocal potentials and tight-binding
Hamiltonians~\cite{graf-prb95}.  For local potentials $m^{-1}_{*}$
becomes the free electron mass --~see \eq{mass-free}~-- and \eq{f-sum}
reduces to the standard form of the $f$-sum rule in crystals.

As it relies on \eq{f-ra-sea}, \eq{f-sum} does not hold for low-energy
continuum models. In the case of graphene, where the low-energy
Hamiltonian is linear in $\k$, the quantity $\langle m_*^{-1} \rangle$
on the right-hand side vanishes by \eq{mass-opt}. Yet, it is well
known that pristine graphene has nonzero optical absorption, which at
the charge-neutrality point is in fact determined by fundamental
constants~\cite{nair2008fine}.  To calculate the integrated absorption
spectrum within the continuum model, one should fall back to
\eqs{f-er}{f-ra-surf} (with a suitable high-frequency cutoff).  At
charge neutrality the Fermi-surface contribution~\eqref{eq:f-ra-surf}
vanishes, and optical absorption is correctly described by the
interband contribution in \eq{f-er}.

The statement that the interband and intraband inverse masses
\eqref{eq:mass-er} and~\eqref{eq:mass-ra} add up to the optical
inverse mass~\eqref{eq:mass-opt} amounts to
\beq
\bra{u\bnk}
\left( \partial^2_\ab H_\k \right)
\ket{u\bnk}
=\partial^2_\ab\enk\ +
2\Re \, \me{\partial_\al u\bnk}{H_\k-\enk}{\partial_\be u\bnk}\,.
\eqlab{ident}
\eeq
When solved for $\partial^2_\ab\enk$, this becomes the effective-mass
theorem, written in a form that remains valid for nonlocal potentials
and low-energy models (including both tight-binding and continuum
Hamiltonians).
To derive \eq{ident} for  nondegenerate energy levels, write
\beq
\begin{aligned}
\partial^2_\ab \enk &=
\partial_\be \me{u\bnk}{(\partial_\al H_\k)}{u\bnk}\\
&= \me{u\bnk}{\left( \partial^2_\ab H_\k \right)}{u\bnk}\\
&+
\me{\partial_\be u\bnk}{(\partial_\al H_\k)}{u\bnk} +
\me{u\bnk}{(\partial_\al H_\k)}{\partial_\be u\bnk}\\
&= \me{u\bnk}{\left( \partial^2_\ab H_\k \right)}{u\bnk}\\
&+
\me{\partial_\be u\bnk}{
\left(\mathbbm{1}-\ket{u\bnk} \bra{u\bnk} \right)(\partial_\al H_\k)}{u\bnk}\\
&+
\me{u\bnk}{(\partial_\al H_\k)
  \left(\mathbbm{1}-\ket{u\bnk} \bra{u\bnk} \right)}{\partial_\be u\bnk}\,,
\end{aligned}
\eeq
and then use the Sternheimer equation~\cite{vanderbilt-book18}
\beq
\left(\mathbbm{1}-\ket{u\bnk} \bra{u\bnk} \right)
\left( \partial_\al H_\k \right) \ket{u\bnk} =
\left( \enk - H_\k \right) \ket{\partial_\al u\bnk}
\,.
\eeq

When the energy levels are degenerate, \eqs{mass-opt}{mass-er} for
$m^{-1}_*(\k)$ and $ m^{-1}_{\rm er}(\k)$ must be generalized by
adding degeneracy indices $dd'$ to the Cartesian indices
$\alpha\beta$.  The transport effective masses are then obtained from
$m^{-1}_{\rm ra}(\k) = m^{-1}_*(\k) - m^{-1}_{\rm er}(\k)$ as detailed
in Ref.~\cite{martins-scipost25}.

\section{Saturated bounds in few-band models}
\seclab{saturated}

The first inequality in \eq{ineqs-2D} relates the imaginary and real
parts of a positive semi-definite matrix $A$ in 2D, and turns into an
equality if and only if $A$ is singular,

\begin{equation}
\det\, A=0 \Leftrightarrow |\mathbf{A}''|=\sqrt{\det\,A'}\,.
\eqlab{singular-2D}
\end{equation}
The same statement applies to the first inequality in \eq{ineqs-3D-a},
where the $2\times2$ matrix $A_z$ is constructed by looking at a
two-dimensional
plane in 3D. Furthermore, the first
inequality in \eq{ineqs-3D-b} results from summing three such
inequalities on orthogonal planes in 3D. Hence,
\begin{equation}
\det\, A_x =
\det\, A_y =
\det\, A_z =
0
\Leftrightarrow\,
|\mathbf{A}''| =
\sqrt{I_2(A')}\,,
\eqlab{singular-3D-I2}
\end{equation}
where $A$ is now a $3\times3$ matrix, and the scalar invariant $I_2$
is defined by \eq{I1-I2-I3}.  In all cases, the (near) saturation of
the aforementioned inequalities relies on an eigenvalue approaching
zero.

Local inequalities with $A=T_p(\mathbf{k})$ saturate in two-band
models.  To see this, consider the ``interband transitions''
expression for $T_p(\k)$ in \eq{T-sos1}, which we repeat here for
convenience,
\begin{equation}
T_p^{\alpha\beta}(\mathbf{k}) =
\sum_{\substack{n\in\Fk\\{l\notin \Fk}}}
\left( \elk-\enk \right)^p\,r^\alpha_{nl\k}\,\left( r^\beta_{nl\k} \right)^*\,.
\eqlab{T-sos1-2}
\end{equation}
Viewed as a matrix in the Cartesian indices, each term is essentially
a projection matrix onto the vector $\r_{nl\k}$, with eigenvalues
$(\epsilon_{l\k}-\epsilon_{n\k})^p\,|\mathbf{r}_{nl\k}|^2$ and
$0$.\footnote{A similar reasoning allows to conclude that the
  Fermi-surface expression in \eq{f-ra-surf} for the integrated
  intraband absorption is locally positive semidefinite in the ground
  state and more generally in thermal equilibrium, since
  $\left( - \partial f/\partial \epsilon \right)_{\epsilon = \enk}$ is
  non-negative.}  Each term therefore contributes a matrix whose rank
is at most one, which is singular in any dimension $d>1$. In two-band
models there is at most one allowed transition at any $\mathbf{k}$,
and $T_p(\k)$ is automatically singular in both 2D and 3D. It follows
that the metric-curvature and mass-moment inequalities saturate
locally in two-band models, as shown in Ref.~\cite{ozawa-prb21} by a
different argument.

A similar reasoning applies to the inequality in \eq{ineqs-3D-c},
which again relates the imaginary and real parts of a positive
semi-definite matrix $A$, which is now $3\times3$. That inequality
comes from the condition $\det A\geq 0$ in \eq{minors}, and hence it
saturates when $A$ is singular,
\beq
\det\, A = 0 \Leftrightarrow \lambda_1 \left( A''_1 \right)^2 +
\lambda_2 \left( A''_2 \right)^2 +
\lambda_3 \left( A''_3 \right)^2
=
\lambda_1 \lambda_2 \lambda_3\,.
\eqlab{ineqs-3D-c-2}
\eeq
When $A=T_p(\mathbf{k})$, this equality holds in any two- or
three-band model. The two-band case was already discussed. With three
bands, there are at most two terms in the sum of \eq{T-sos1-2}; since
$\text{rank}(A+B) \leq \text{rank}(A) + \text{rank}(B)$, it follows
that the rank of $T_p(\k)$ is at most two, making it singular in 3D.

Finally, the real part of $T_p(\k)$ is also singular for two-band
models in 3D: since
\beq
T_p'(\k) =
\frac{1}{2}\left[ T_p(\k) +
T_p^\dagger(\k) \right]
\eeq
is the sum of two projections, it has at most rank two, becoming
singular when $d>2$. Thus, in 3D two-band models the quantum metric
and the inverse interband effective mass have a null eigenvalue at
every $\k$. We have verified this numerically for the layered Haldane
model.

Consider now global sum-rule inequalities, whose saturation was
discussed in \sref{saturation} from the perspective of optical
absorption.  Mathematically, it again boils down to how close the
matrix $S_p \propto \int_\k T_p(\k)$ of \eq{sum-rule} is to being
singular.  Contributions are now included not only from multiple bands
as in \eq{T-sos1-2}, but also from every $\k$, making it a nontrivial
condition even for few-band models. If $T_p(\k)$ is uniform and
singular over the entire BZ, the global and local inequalities are the
same and both saturate, as seen in Landau levels~\cite{ozawa-prb21}.

Flat bands with a uniform and singular metric-curvature tensor
$T_0(\k)$ across the BZ have recently attracted considerable interest,
as they are suitable for hosting fractional Chern insulating
states~\cite{roy-prb14}. A near-zero $\det T_0(\k)$ at every $\k$ has
been found in some topological flat-band models, such as that of
\sref{flatband-magnetization}. Current research explores saturation of
the global metric-curvature inequality beyond uniform metric and the
standard Landau-level
structure~\cite{ledwith-prb23,fujimoto-prl25,liu-prx25}.

\section{$T_p(\k)$ tensor  for Landau levels at integer filling}
\seclab{LL}

The simplest example of a system with broken time-reversal symmetry
and topologically nontrivial bands is a 2D electron gas in a uniform
transverse magnetic field ${\bf B} = B\hat{\bf z}$.  In the Landau
gauge, the Hamiltonian reads
\begin{equation}
H_\mathbf{k} =
\frac{\hbar^2}{2\,m_{\rm e}}
\left[ k_x^2 + \left( k_y-\text{sign}(B)\,x/l_B^2 \right)^2 \right]\,,
\eqlab{SHO}
\end{equation}
with $l_B=\sqrt{\hbar/|e B|}$ the magnetic length.  Magnetic
translation symmetry permits the construction of a square Bravais
lattice with a flux quantum $h/|e|$ threading the primitive cell.  Its
area is $A_{\rm c}= h/|eB|$, and that of the BZ is
$(2\pi)^2/A_{\rm c}=2\pi/l_B^2$.  The flat-band spectrum,
$\epsilon_{\mathbf{k}n}=(n+1/2)\hbar\omega_{\rm c}$, has a fixed
energy separation set by the cyclotron frequency
$\omega_{\rm c} = |e B|/m_{\rm e}$.

To evaluate the tensor $T_p(\k)$ for the group of $\nu$ lowest Landau
levels, insert in \eq{T-sos1} the expression given in
Ref.~\cite{ozawa-prb21} for the interband dipole matrix elements
$\mathbf{r}_{ln\mathbf{k}}$. The result is independent of $\k$ and
reads
\begin{equation}
T_p =
\frac{\nu}{2}\,l_B^2\;(\hbar\omega_c)^p
\begin{pmatrix}
1 & i\,\text{sgn}(B) \\
-i\,\text{sgn}(B) & 1
\end{pmatrix}\,.
\eqlab{Tp-LLL}
\end{equation}
Plugging the Berry curvature $\Omega = -2T''_0$ into \eq{C} gives
$C=-\nu\, \text{sgn}(B)$~\cite{ozawa-prb21}.

All applicable inequalities saturate for low-lying Landau levels at
integer filling~\cite{ozawa-prb21,onishi-prx24,shinada-prb25}; as
there is no $\k$ dependence, local and global relations are
identical. Take \eq{local-2D} with $A=T_p$: since $T'_p$ is isotropic,
its right inequality saturates; and since diagonal and off-diagonal
entries in \eq{Tp-LLL} have the same magnitude, the left inequality
saturates as well.  The gap and Cauchy-Schwarz inequalities of
\sref{spectral} saturate because the absorption spectrum is
concentrated at the gap frequency $\Eg = \hbar \w_{\rm c}$; this
follows from dipole selection rules, which only allow transitions
between neighboring Landau levels~\cite{ozawa-prb21}.  All remaining
relations are obtained through combinations of the ones above, and
hence they saturate as well.

The above considerations apply when the $\nu$ lowest Landau levels are
full.  Suppose, for discussion purposes, that a single high-lying
Landau level $\nu > 0$ is full. Instead of \eq{Tp-LLL}, we now have,
for $B>0$,
\begin{equation}
T_p=\,l_B^2\,(\hbar\omega_c)^p\,\frac{\nu+1}{2}
\begin{pmatrix}
1 & i \\
-i & 1
\end{pmatrix}
+l_B^2\,(-\hbar\omega_c)^p\,\frac{\nu}{2}
\begin{pmatrix}
1 & -i \\
i & 1
\end{pmatrix}\,.
\eqlab{Tp-LL}
\end{equation}
The first (second) term comes from transitions to level $\nu+1$
($\nu-1$). If $p$ is odd, one is positive-semidefinite and the
other is negative-semidefinite. Their sum is neither,
\begin{equation}
T_p=l_B^2\,(\hbar\omega_c)^p
\begin{pmatrix}
\frac{1}{2} & i\left( \nu + \frac{1}{2} \right) \\
-i\left( \nu + \frac{1}{2} \right) & \frac{1}{2}
\end{pmatrix}\qquad\text{(for $p$ odd)}\,.
\eqlab{Tp-LL-2}
\end{equation}
With $p=1$ (mass-moment tensor), this leads to
\beq
\left| \langle \mm \rangle \right| =
\left(\nu + \frac{1}{2} \right)
\tr\left< \mu_{\rm B}^{\rm er}\right> =
(2\nu+1)\mu_{\rm B}\,,
\eeq
which violates the bounds in \eq{ineq-m2} for any $\nu > 0$.

\end{appendix}

\let\k\ksave
\let\r\rsave
\let\a\asave
\let\b\bsave

\bibliography{bib.bib}

\end{document}